\documentclass[journal]{IEEEtran}
\usepackage{amsmath,amssymb,amsthm} 
\usepackage{mathrsfs}
\usepackage{algorithm}  
\usepackage{algpseudocode}

\usepackage{graphicx}
\graphicspath{{figures/}}

\makeatletter
\@ifclasswith{IEEEtran}{compsoc}{%
	\usepackage[caption=false,font=normalsize]{subfig} 
}{%
	\usepackage[caption=false,font=footnotesize]{subfig} 
}

\usepackage{etoolbox}    

\usepackage{stfloats}
\AtBeginDocument{
	\addtolength{\abovedisplayskip}{-1ex}
	\addtolength{\belowdisplayskip}{-1ex}
	\addtolength{\abovedisplayshortskip}{-1ex}
	\addtolength{\belowdisplayshortskip}{-1ex}
}

{\tiny }

\usepackage{enumitem}
\setlist{nosep,left=0pt}

\makeatletter
\patchcmd{\algorithmic}{\itemsep\z@}{\itemsep2pt}{}{}
\makeatother

\usepackage{xcolor}  
\usepackage{bm}
\usepackage{cite}
\usepackage[colorlinks,
urlcolor=blue,
linkcolor=blue,
anchorcolor=blue,
citecolor=blue]{hyperref}

\usepackage{longtable}
\usepackage{multirow}
\usepackage{tabularx}

\usepackage{lettrine}

\begin{document}
	\title{Design of a Speech Enhancement-Based Deep Semantic Autoencoder for Wireless Communications}
	
	
	\author{
		Xiaoling Han,~\IEEEmembership{Student Member,~IEEE,}
		Bin Lin*,~\IEEEmembership{Senior Member,~IEEE,}
		Nan Wu,~\IEEEmembership{Member,~IEEE,}
		Tony Q. S. Quek,~\IEEEmembership{Fellow,~IEEE}
		\thanks{Manuscript received XXX; revised XXX; accepted XXX. Date of publication XXX; date of current version XXX. 
		}
		
		\thanks{
			The work was supported by the National Natural Science Foundation of China (No. 62371085) and Fundamental Research Funds for the Central Universities (No. 3132023514). This paper is supported in part by the National Research Foundation, Singapore and Infocomm Media Development Authority under its Future Communications Research \& Development Programme. This paper was supported by the China Scholarship Council (CSC).
			(Corresponding author Bin Lin.)
		}
		
		\thanks{
			Xiaoling Han, Bin Lin and Nan Wu are with the Department of Information Science and Technology of Dalian Maritime University, Dalian, China, 116026 (email: xiaolinghan@dlmu.edu.cn, binlin@dlmu.edu.cn, wu.nan@dlmu.edu.cn). 
			Tony Q. S. Quek is with the Information Systems Technology and Design Pillar, Singapore University of Technology and Design, Singapore, 487372 (email: tonyquek@sutd.edu.sg).
		}
		
		\thanks{
			This manuscript has been accepted by IEEE Transactions on Vehicular Technology, DOI: 10.1109/TVT.2026.3658080.
		}
		
	}
	
	\markboth{Journal of \LaTeX\ Class Files,~Vol.~14, No.~8, August~2015}%
	{Shell \MakeLowercase{\textit{\textit{et al.}}} Bare Demo of IEEEtran.cls for IEEE Journals}
	
	\maketitle
	
	\begin{abstract}
		Speech communication plays a crucial role in the Internet of Things and autonomous networks. However, traditional communication systems suffer from insufficient semantic extraction and limited noise suppression capability. To address these issues, we propose a Speech enhancement-based Deep Semantic Autoencoder (S-DSA) communication system for wireless communications. It consists of a dual-branch semantic encoder, a channel encoder, a channel decoder, and two semantic decoders, enabling joint source–channel coding for robust transmission. A key novelty lies in an adaptive semantic noise decoupling mechanism that disentangles target speech semantics from noise representations through an interaction module, thus maintaining semantic consistency under low SNR conditions. Furthermore, a joint training strategy optimizes speech reconstruction and noise estimation, while a lightweight testing strategy reduces complexity by disabling the noise-decoding path. The proposed S-DSA enables more effective separation between speech and noise, and yields higher robustness as well as better reconstruction quality. These findings highlight the potential of the proposed S-DSA as a practical solution for next-generation intelligent wireless communications. 
	\end{abstract}
	\vspace{-2 mm}
	
	\IEEEpeerreviewmaketitle
	
	\vspace{-1 mm}
	\section{Introduction}
	\vspace{-1 mm}
	\lettrine[lines=2, loversize=0.1]{I}{n} emerging technological paradigms such as the Internet of Things (IoT), communication systems face increasingly stringent requirements due to dynamic environments and heterogeneous service demands \cite{YeOJVT2021,YeTVT2018}. In these scenarios, speech communication plays a particularly critical role as a key enabler of intelligent coordination. In large-scale networks, guaranteeing End-to-End (E2E) quality of service remains a significant challenge \cite{YeMVT2018}. However, traditional methodologies have inherent limitations. Their waveform-centric design, grounded in Shannon’s information theory, primarily ensures reliable  syntactic transmission, while their fixed signal processing frameworks lack adaptability in noisy environments. Therefore, signal distortion and semantic loss become severe in low-SNR scenarios \cite{zhang2022theory}. These drawbacks highlight the urgent need for new communication paradigms to improve communication efficiency.
	
	Recently, semantic communication, as an emerging paradigm, is promising for 6G and beyond \cite{tiaozhan}. Many seminal studies have laid the theoretical foundations for this transformation. Qin \textit{et al.} \cite{QinBits2Se} reviewed how Artificial Intelligence (AI) and Deep Learning (DL) enable communication systems to evolve from bits to semantics, highlighting semantic communication frameworks and resource optimization. Additionally, Getu \textit{et al.} \cite{GetuSurvey} surveyed the semantic communication landscape in the context of 6G, introducing semantic information measures and use cases, and outlining key challenges such as theoretical foundations and system integration. Lu \textit{et al.} \cite{endecoder} re-examined modern communication from semantic coding to semantic transmission, providing theoretical insights and technical directions. Wheeler and Natarajan \cite{EngineeringSemantic} offered a systematic survey on engineering aspects of semantic communication. Gündüz \textit{et al.} \cite{Beyondbits} discussed task-oriented communication that exploits context and semantics to achieve communication goals. Together, these works establish the foundations of semantic communication and inspire its application across diverse modalities, including text, images, speech, and so on.
	
	Among these application scenarios, speech communication is critical since it is the primary medium of human interaction and directly impacts the quality of human–machine coordination. Numerous studies have concentrated on the encoding and transmission in speech communication scenarios. For example, Liu \textit{et al.} \cite{speechtrad} proposed a DL-based method that improved the way of speech communication quality assessment. It used large-scale invasive simulation data to improve the accuracy and generalization of non-invasive speech quality assessment. Dong \textit{et al.} \cite{speechtrad2} proposed EDSep, a novel single-channel method based on fractional matching of stochastic differential equations. It enhanced the generative modeling of speech source separation by optimizing the training and sampling efficiency. Nevertheless, diverse tasks under noisy environments demand robust communication links, whereas waveform-based speech methods often fail to preserve semantic content \cite{Niazmand2025}.
	
	Speech semantic communication has therefore attracted increasing attention. Several pioneering works have demonstrated its potential. Qin and Weng \cite{DeepSC-S} designed the DeepSC-S, a semantic communication system. Based on the attention mechanism and squeeze-and-excitation networks, it improved the accuracy of speech signal recovery. Tan \textit{et al.} \cite{speechyuyi1} proposed the emotional information-assisted semantic communication framework EESC-S for speech transmission. It jointly extracted the content semantics and emotional semantics of speech, improving the speech transmission performance. Yang \textit{et al.} \cite{Deep-STS} designed the Deep-STS system for low-bandwidth speech communication. By extracting the text information and timbre features in speech, it achieved ultra-low bit rate communication. Zhou \textit{et al.} \cite{DeepSC-TS} proposed the DeepSC-TS, an E2E speech semantic communication system based on transformer, which effectively eliminated semantic signal noise and showed good performance in different channel environments. Chen \textit{et al.} \cite{speechyuyi2} proposed a perception-motivated, low-complexity speech semantic communication method. It improved the transmission efficiency and ensured that the reconstructed speech is consistent with human auditory perception.
	
	Beyond speech transmission, many works have applied semantic communication to other tasks. Han \textit{et al.} \cite {S2T1} constructed a semantic-oriented speech transmission system that transmitted semantic-related information and reconstructed speech. Its E2E transceiver effectively improved the accuracy of speech-to-text transmission and the quality of speech reconstruction. Weng \textit{et al.} \cite {S2T2} proposed the SAC-ST system for Multiple-Input Multiple-Output communication scenarios, which compresses semantic information and ensures high-fidelity transmission. It also reduced the dependence on channel state information through the channel estimation network, showing significant advantages under low Signal-to-Noise Ratios (SNR). Weng \textit{et al.} \cite{qinDeepSC-ST} developed DeepSC-ST for low-bandwidth scenarios, which integrates speech recognition and synthesis. This approach reduces data transmission and outperforms traditional systems in dynamic channels. These studies have advanced speech semantic communication in terms of accuracy, expressiveness, bitrate reduction, and robustness. Recent work \cite{JiaVTC2024} further illustrates the challenges of highly dynamic environments, where environmental noise, i.e., non-speech background interference with time-varying statistics, can severely degrade semantic preservation. However, existing approaches generally lack mechanisms for explicit semantic noise decoupling in noisy environments.
	
	Based on the above analysis, we propose a Speech enhancement-based Deep Semantic Autoencoder (S-DSA)  communication system for wireless communications. 
	Leveraging DL, the proposed S-DSA adopts an E2E Joint Source–Channel Coding (JSCC) framework to decouple semantics from noise and jointly enhance, extract, and transmit speech semantic representations.
	This design enables robust speech transmission in noisy environments, improving resilience to interference and reconstruction fidelity. The main contributions of this paper are as follows. 
	
	\begin{enumerate}
		
		\item We propose a semantic communication system, named S-DSA. It is specifically designed for communication in noisy environments. It consists of a dual-branch semantic encoder, a channel encoder, a channel decoder, and two semantic decoders. By employing JSCC, the proposed S-DSA enables high-fidelity transmission of target speech features and restores the semantics.
		
		\item We design an adaptive semantic noise decoupling mechanism. The dual-branch semantic encoder dynamically refines speech semantic features and noise statistical features to facilitate feature-level decoupling from the mixed signal. The proposed S-DSA can simultaneously extract the target speech and estimate the environmental noise. 
		
		\item We develop training and testing strategies. The training algorithm adopts an E2E joint optimization strategy to learn network parameters effectively. By using gradient descent, the weighted sum of the speech reconstruction loss and the noise estimation loss is minimized. The testing algorithm disables the noise decoding branch to form a lightweight inference path and reduce computational overhead. It evaluates the performance of speech enhancement and reconstruction under the simulation setting.
		
		\item We conduct extensive simulation experiments. Results show that, compared with benchmarks, the proposed S-DSA operates effectively under different environmental noise conditions and transmission channels. It not only effectively separates target speech from environmental noise, but also supports the transmission and reconstruction of target speech with reduced interference. In addition, the architecture reduces computational overhead, providing a cost-effective solution for intelligent communication with limited resources.
		
	\end{enumerate}
	
	The remainder of this paper is organized as follows. Section II describes the problem and formulates the system model. Section III develops the training and testing strategy. Section IV reports experimental results across different settings. Section V concludes this paper.

	\begin{figure*}[t]
		\centering
		\subfloat[Modular overview\label{architecture}]%
		{\includegraphics[width=1\textwidth]{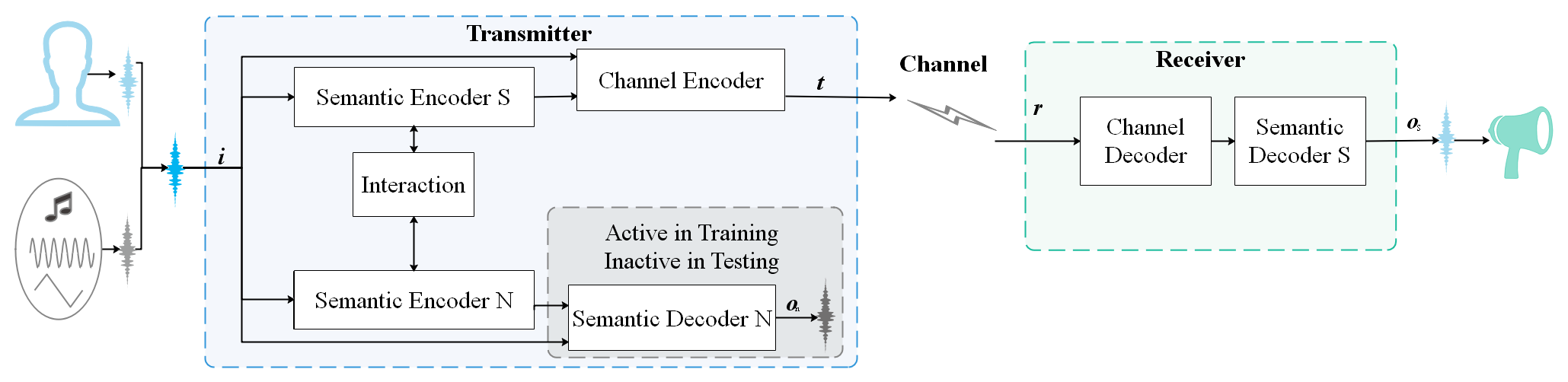}}\par
		\subfloat[Detailed design\label{structure}]%
		{\includegraphics[width=1\textwidth]{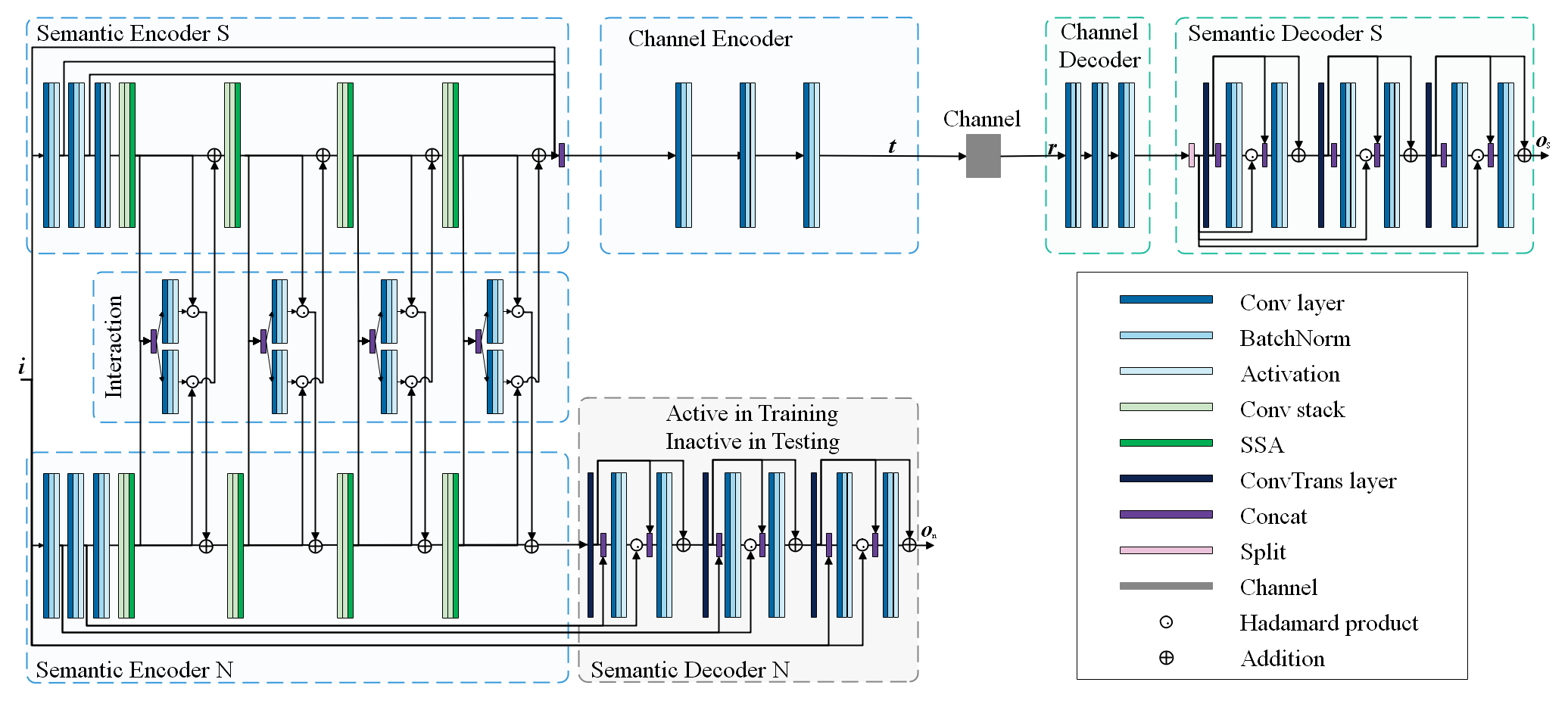}}
		\caption{Architecture of the proposed S-DSA.} 
		\label{model}
	\end{figure*}

	\vspace{-1 mm}
	\section{System Model}
	\vspace{-1 mm}
	This section describes the problem addressed, introduces the model design rationale, formulates the model, and states the optimization objective.

    To decouple semantic noise and reliably transmit target speech, we propose S-DSA, an end to end JSCC semantic system integrating speech enhancement with semantic encoding and decoding.
    The transmitter adopts a modular hierarchical design, including a dual-branch semantic encoder, a semantic decoder N, and a channel encoder, enabling mixed speech separation and efficient transmission of the target speech. The receiver comprises a channel decoder and semantic decoder S, which reconstructs the target speech.  
	By employing JSCC, the proposed S-DSA enables high-fidelity recovery of the target semantics. 	
	The overall architecture is shown in Fig. \ref{model}. In particular, Fig. \ref{architecture} presents the modular architecture, whereas Fig. \ref{structure} depicts the detailed structural design.
	The corresponding structural parameters are summarized in Tab. \ref{table1}. 
	Specifically, $B$, $T$, and $F$ denote the batch size, number of frames, and number of frequency bins, respectively.
	$C_{\rm in}$ is the input dimension, $4D_{\rm s}$ is the semantic feature dimension, and $D_{\rm c}$ is the channel transmitted dimension.

	We consider the input signal in the time domain, where the clean target speech $\bm{i}_{\rm s}$ and the additive environmental noise $\bm{i}_{\rm n}$  are combined to form the observed mixture. This process can be expressed as
	\begin{equation}
		\bm{i} = \bm{i}_{\rm s} + \bm{i}_{\rm n}, \quad 
		\bm{i}_{\rm s} \sim p(\bm{i}_{\rm s}), \; 
		\bm{i}_{\rm n} \sim \mathcal{N}(0,\sigma_n^2),
	\end{equation}
	where $p(\bm{i}_{\rm s})$ represents the prior distribution of $\bm{i}_{\rm s}$, and $\mathcal{N}(\bm{0},\sigma_n^2\bm{I})$ denotes zero-mean Additive White Gaussian Noise (AWGN) with per-entry variance $\sigma_n^2$. $\bm{i}_{\rm s}$ and $\bm{i}_{\rm n}$ are assumed to be independent. 
	The time domain mixture $\bm{i}$ is converted into the time frequency domain using the Short Time Fourier Transform (STFT). The resulting complex spectrogram can be expressed as
	\begin{equation}
		\bm{Y} = \mathcal{S}(\bm{i}) \in \mathbb{R}^{B\times T\times F},
	\end{equation}
	where $\mathcal{S}(\cdot)$ denotes the mapping function of STFT. For network input, we adopt a two channel representation using the real and imaginary parts, which can be expressed as
	\begin{equation}
		\bm{I} = \big[\Re(\bm{Y})\ \|\ \Im(\bm{Y})\big]\in\mathbb{R}^{B\times C_{\rm in}\times T\times F}, \quad C_{\rm in}=2,
	\end{equation}
	where $\Re(\cdot)$ and $\Im(\cdot)$ denote the real and imaginary parts of $\bm{Y}$, respectively. Similarly, we define $\bm{S}$ and $\bm{N}$ as the complex spectrograms of $\bm{i}_{\rm s}$ and $\bm{i}_{\rm n}$, respectively.
	
	\begin{table*}
		\centering
		\caption{Architectural parameters of the proposed S-DSA.}\label{table1}
		\renewcommand{\arraystretch}{1.05}
		\begin{tabularx}{1\textwidth}{@{}p{2.4cm} p{4.5cm} p{6cm} X@{ }}
			\hline
			\textbf{End} & \textbf{Layer Category} & \textbf{Layer Name} & \textbf{Output Dimension} \\
			\hline
			
			\multirow{31}{*}{Transmitter}
			& \multirow{19}{*}{%
				\begin{tabular}[t]{@{}l@{}}
					Dual-branch semantic encoder\\
					\footnotesize \\
					\footnotesize Only the semantic encoder S is shown; 
					\\
					\footnotesize the semantic encoder N is identical.
				\end{tabular}
			}
			& Conv block (Semantic encoder S) & [$B$, $D_{\rm s}$, $T$, $F$] \\
			& & Conv block (Semantic encoder S) & [$B$, $2D_{\rm s}$, $T$, $F/2$] \\
			& & Conv block (Semantic encoder S) & [$B$, $4D_{\rm s}$, $T$, $F/4$] \\

			& & Conv stack (Semantic encoder S) & [$B$, $4D_{\rm s}$, $T$, $F/4$] \\
			& & Conv stack (Semantic encoder S) & [$B$, $T$, $F/4$, $4D_{\rm s}$] \\
			& & SSA (Semantic encoder S) & [$B$, $T$, $F/4$, $4D_{\rm s}$] \\
			
			& & Interaction block 1 & [$B$, $4D_{\rm s}$, $T$, $F/4$] \\

			& & Conv stack (Semantic encoder S) & [$B$, $4D_{\rm s}$, $T$, $F/4$] \\
			& & Conv stack (Semantic encoder S) & [$B$, $T$, $F/4$, $4D_{\rm s}$] \\
			& & SSA (Semantic encoder S) & [$B$, $T$, $F/4$, $4D_{\rm s}$] \\
			
			& & Interaction block 2 & [$B$, $4D_{\rm s}$, $T$, $F/4$] \\

			& & Conv stack (Semantic encoder S) & [$B$, $4D_{\rm s}$, $T$, $F/4$] \\
			& & Conv stack (Semantic encoder S) & [$B$, $T$, $F/4$, $4D_{\rm s}$] \\
			& & SSA (Semantic encoder S) & [$B$, $T$, $F/4$, $4D_{\rm s}$] \\
			
			& & Interaction block 3 & [$B$, $4D_{\rm s}$, $T$, $F/4$] \\

			& & Conv stack (Semantic encoder S) & [$B$, $4D_{\rm s}$, $T$, $F/4$] \\
			& & Conv stack (Semantic encoder S) & [$B$, $T$, $F/4$, $4D_{\rm s}$] \\
			& & SSA (Semantic encoder S) & [$B$, $T$, $F/4$, $4D_{\rm s}$] \\
			& & Interaction block 4 & [$B$, $4D_{\rm s}$, $T$, $F/4$] \\
			\cline{2-4}
			
			& \multirow{9}{*}{Semantic decoder N}
			& ConvTrans layer & [$B$, $2D_{\rm s}$, $T$, $F/2$] \\
			& & Conv block & [$B$, $2D_{\rm s}$, $T$, $F/2$] \\
			& & Conv block & [$B$, $2D_{\rm s}$, $T$, $F/2$] \\
			& & ConvTrans layer & [$B$, $D_{\rm s}$, $T$, $F$] \\
			& & Conv block & [$B$, $D_{\rm s}$, $T$, $F$] \\
			& & Conv block & [$B$, $D_{\rm s}$, $T$, $F$] \\
			& & ConvTrans layer & [$B$, $C_{\rm in}$, $T$, $F$] \\
			& & Conv block & [$B$, $C_{\rm in}$, $T$, $F$] \\
			& & Conv block & [$B$, $C_{\rm in}$, $T$, $F$] \\
			\cline{2-4}
			
			& \multirow{3}{*}{Channel encoder}
			& Conv block & [$B$, $4D_{\rm c}$, $T$, $F/4$] \\
			& & Conv block & [$B$, $2D_{\rm c}$, $T$, $F/4$] \\
			& & Conv block & [$B$, $D_{\rm c}$, $T$, $F/4$] \\
			\hline
			
			\multirow{12}{*}{Receiver}
			& \multirow{3}{*}{Channel decoder}
			& Conv block & [$B$, $2D_{\rm c}$, $T$, $F/4$] \\
			& & Conv block & [$B$, $4D_{\rm c}$, $T$, $F/4$] \\
			& & Conv block & [$B$, $7D_{\rm s}+C_{\rm in}$, $T$, $F$] \\
			\cline{2-4}
			
			& \multirow{9}{*}{Semantic decoder S}
			& ConvTrans layer & [$B$, $2D_{\rm s}$, $T$, $F/2$] \\
			& & Conv block & [$B$, $2D_{\rm s}$, $T$, $F/2$] \\
			& & Conv block & [$B$, $2D_{\rm s}$, $T$, $F/2$] \\
			& & ConvTrans layer & [$B$, $D_{\rm s}$, $T$, $F$] \\
			& & Conv block & [$B$, $D_{\rm s}$, $T$, $F$] \\
			& & Conv block & [$B$, $D_{\rm s}$, $T$, $F$] \\
			& & ConvTrans layer & [$B$, $C_{\rm in}$, $T$, $F$] \\
			& & Conv block & [$B$, $C_{\rm in}$, $T$, $F$] \\
			& & Conv block & [$B$, $C_{\rm in}$, $T$, $F$] \\
			\hline
		\end{tabularx}
	\end{table*}
	
	\vspace{-2 mm}
	\subsection{Dual-branch semantic encoder} 
	\vspace{-1 mm}
	The dual-branch semantic encoder is designed to promote orthogonality between speech and noise features via an adaptive semantic noise decoupling mechanism. This mechanism explicitly models the distinct characteristics of speech and noise, encouraging their separation in the feature space. It comprises semantic encoder S, semantic encoder N, and the interaction module. Each semantic encoder contains three convolutional (Conv) blocks followed by a stack of four Residual–Attention (RA) blocks. The interaction module consists of four interaction blocks, which is a coequal component that is invoked after each RA block.
	
	The two semantic encoders operate in parallel. Semantic encoder S extracts and encodes target speech semantics, while semantic encoder N estimates and encodes environmental noise statistics. Each Conv block consists of a Conv layer, batch normalization (BatchNorm), and a ReLU activation. Given the input $\bm{I}$, after processing through three Conv blocks in each branch, the feature maps can be expressed as
	\begin{align}
		\bm{I}^{\rm c1}_{\rm s} &= \mathcal{E}^{\rm {s1}}_{\rm{en}}(\bm{I}), &
		\bm{I}^{\rm c2}_{\rm s} &= \mathcal{E}^{\rm {s2}}_{\rm{en}}(\bm{I}^{\rm c1}_{\rm s}), &
		\bm{I}^{\rm c3}_{\rm s} &= \mathcal{E}^{\rm {s3}}_{\rm{en}}(\bm{I}^{\rm c2}_{\rm s}), \\
		\bm{I}^{\rm c1}_{\rm n} &= \mathcal{E}^{\rm {n1}}_{\rm{en}}(\bm{I}), &
		\bm{I}^{\rm c2}_{\rm n} &= \mathcal{E}^{\rm {n2}}_{\rm{en}}(\bm{I}^{\rm c1}_{\rm n}), &
		\bm{I}^{\rm c3}_{\rm n} &= \mathcal{E}^{\rm {n3}}_{\rm{en}}(\bm{I}^{\rm c2}_{\rm n}),
	\end{align}
	where $\mathcal{E}^{(\cdot j)}_{\rm{en}}$ denotes the mapping function of $j$-th Conv block, $\bm{I}^{\rm c1}_{\rm s}, \bm{I}^{\rm c1}_{\rm n}\in\mathbb{R}^{B\times D_{\rm s}\times T\times F}$,
	$\bm{I}^{\rm c2}_{\rm s}, \bm{I}^{\rm c2}_{\rm n}\in\mathbb{R}^{B\times 2D_{\rm s}\times T\times (F/2)}$,
	and $\bm{I}^{\rm c3}_{\rm s}, \bm{I}^{\rm c3}_{\rm n}\in\mathbb{R}^{B\times 4D_{\rm s}\times T\times (F/4)}$.

	Each RA block consists of two residual Conv stacks and one Separable Self-Attention (SSA). Starting from $\bm{I}^{\rm c3}_{\rm s}$ and $\bm{I}^{\rm c3}_{\rm n}$, each branch applies four RA blocks. For stage $k\!=\!1,\dots,4$, the output of the RA can be expressed as
	\begin{align}
		\bm{I}^{{\rm r}k}_{\rm s}=\mathrm{RA}^{k}_{\rm s}\!\big(\hat{\bm{I}}^{{\rm r}(k-1)}_{\rm s}\big),\qquad
		\bm{I}^{{\rm r}k}_{\rm n}=\mathrm{RA}^{k}_{\rm n}\!\big(\hat{\bm{I}}^{{\rm r}(k-1)}_{\rm n}\big),
	\end{align}
	with initialization $\hat{\bm{I}}^{{\rm r}0}_{\rm s}=\bm{I}^{\rm c3}_{\rm s}$ and $\hat{\bm{I}}^{{\rm r}0}_{\rm n}=\bm{I}^{\rm c3}_{\rm n}$.
	Within each RA block, two residual Conv stacks are applied in sequence. Every Conv stack consists of Conv layer, BatchNorm, ReLU, Conv layer, BatchNorm. After processing through two Conv stacks in each branch, the output can be expressed as
	\begin{align}
		\bm{I}^{{\rm r}k,\,{\rm res}1}_{\rm s} &= \mathcal{R}^{\rm s}_{k,1}\!\big(\hat{\bm{I}}^{{\rm r}(k-1)}_{\rm s}\big), &
		\bm{I}^{{\rm r}k,\,{\rm res}2}_{\rm s} &= \mathcal{R}^{\rm s}_{k,2}\!\big(\bm{I}^{{\rm r}k,\,{\rm res}1}_{\rm s}\big),\\
		\bm{I}^{{\rm r}k,\,{\rm res}1}_{\rm n} &= \mathcal{R}^{\rm n}_{k,1}\!\big(\hat{\bm{I}}^{{\rm r}(k-1)}_{\rm n}\big), &
		\bm{I}^{{\rm r}k,\,{\rm res}2}_{\rm n} &= \mathcal{R}^{\rm n}_{k,2}\!\big(\bm{I}^{{\rm r}k,\,{\rm res}1}_{\rm n}\big),
	\end{align}
	where $\bm{I}^{{\rm r}k,\,{\rm res}1}_{\rm s}, \bm{I}^{{\rm r}k,\,{\rm res}2}_{\rm s}, \bm{I}^{{\rm r}k,\,{\rm res}1}_{\rm n}, \bm{I}^{{\rm r}k,\,{\rm res}2}_{\rm n} \in\mathbb{R}^{B\times 4D_{\rm s} \times T \times (F/4)}$. The residual outputs in channel-last can be expressed as
	\begin{align}
		\bm{F}^{(k)}_{{\rm res},\,{\rm s}}\triangleq & \bm{I}^{{\rm r}k,{\rm res}2}_{\rm s}\in\mathbb{R}^{B\times T\times (F/4)\times 4D_{\rm s}},\qquad \\
		\bm{F}^{(k)}_{{\rm res},\,{\rm n}}\triangleq & \bm{I}^{{\rm r}k,{\rm res}2}_{\rm n}\in\mathbb{R}^{B\times T\times (F/4)\times 4D_{\rm s}}.
	\end{align}
	Then, SSA models long-range dependencies along time and frequency. First compress channels by $\psi(\cdot)$ from $4D_{\rm s}$ to $D_{\rm s}$ and build temporal self-attention, which can be expressed as
	\begin{align}
		\bm{Z}^{(k)}_{\rm s}&=\psi\!\big(\bm{F}^{(k)}_{{\rm res},\,{\rm s}}\big)\in\mathbb{R}^{B\times T\times (F/4)\times D_{\rm s}},
		\\
		\bm{Z}^{(k)}_{{\rm s},t}&\in\mathbb{R}^{B\times T\times d_t},\quad d_t=D_{\rm s}\cdot(F/4),
		\\
		\bm{A}^{(k)}_{{\rm s},t}&=\mathrm{softmax}\!\Big(\frac{\bm{Z}^{(k)}_{{\rm s},t}\,\bm{Z}^{(k)\top}_{{\rm s},t}}{\sqrt{d_t}}\Big)\in\mathbb{R}^{B\times T\times T},
		\\
		\bm{F}^{(k)}_{{\rm temp},\,{\rm s}}&=\bm{F}^{(k)}_{{\rm res},\,{\rm s}}+\varphi\!\big(\mathrm{unflatten}_t(\bm{A}^{(k)}_{{\rm s},t}\bm{Z}^{(k)}_{{\rm s},t})\big),
	\end{align}
	where $\varphi(\cdot)$ restores the channel dimension from $D_{\rm s}$ back to $4D_{\rm s}$. Similarly, build spectral self-attention by reshaping along $T$, which can be expressed as 
	\begin{align}
		\bm{Z}^{(k)}_{{\rm s},f}&\in\mathbb{R}^{B\times (F/4)\times d_f},\quad d_f=D_{\rm s}T,
		\\
		\bm{A}^{(k)}_{{\rm s},f}&=\mathrm{softmax}\!\Big(\frac{\bm{Z}^{(k)}_{{\rm s},f}\,\bm{Z}^{(k)\top}_{{\rm s},f}}{\sqrt{d_f}}\Big)\in\mathbb{R}^{B\times (F/4)\times (F/4)},
		\\
		\bm{F}^{(k)}_{{\rm freq},\,{\rm s}}&=\bm{F}^{(k)}_{{\rm res},\,{\rm s}}+\varphi\!\big(\mathrm{unflatten}_f(\bm{A}^{(k)}_{{\rm s},f}\bm{Z}^{(k)}_{{\rm s},f})\big).
	\end{align}
	The residual, temporal, and frequency streams are concatenated and projected by a $1{\times}1$ head to obtain the RA output, which can be expressed as
	\begin{align}
		\bm{F}^{(k)}_{{\rm cat},\,{\rm s}}&=\big[\bm{F}^{(k)}_{{\rm res},\,{\rm s}}\ \|\ \bm{F}^{(k)}_{{\rm temp},\,{\rm s}}\ \|\ \bm{F}^{(k)}_{{\rm freq},\,{\rm s}}\big]\in\mathbb{R}^{B\times T\times (F/4)\times 3\cdot 4D_{\rm s}},
		\\
		\bm{I}^{\rm r k}_{\rm s}&=\omega^{(k)}_{\rm s}\!\big(\bm{F}^{(k)}_{{\rm cat},\,{\rm s}}\big)\in\mathbb{R}^{B\times T\times (F/4)\times 4D_{\rm s}},
	\end{align}
	where $\omega^{(k)}_{\rm s/n}(\cdot)$ is a $1{\times}1$ fusion projection restoring the channel dimension to $4D_{\rm s}$. For the noise branch, the operations are analogous, $\bm{I}^{{\rm r}k}_{\rm n}\in\mathbb{R}^{B\times T\times (F/4)\times 4D_{\rm s}}$.

	The interaction module performs cross-gated message passing between the speech and noise branches. A joint context is formed by concatenating the post-RA features, and two independent Conv projections produce element-wise gates that modulate cross-branch messages. Given the features $\bm{I}^{{\rm r}k}_{\rm s},\bm{I}^{{\rm r}k}_{\rm n}\in\mathbb{R}^{B\times T\times (F/4)\times 4D_{\rm s}}$ for the speech and noise branches, respectively, first form a joint context by channel-wise concatenation, which can be expressed as
	\begin{equation}
		\bm{U}^{(k)}=\big[\bm{I}^{{\rm r}k}_{\rm n}\ \|\ \bm{I}^{{\rm r}k}_{\rm s}\big]\in\mathbb{R}^{B\times T\times (F/4)\times 2\cdot 4D_{\rm s}}.
	\end{equation}
	Two independent projections generate the element-wise gates, which can be expressed as
	\begin{equation}
		\bm{E}_{n\rightarrow s}^{(k)} = \sigma\!\big(g_{n\rightarrow s}^{(k)}(\bm{U}^{(k)})\big),\qquad
		\bm{E}_{s\rightarrow n}^{(k)} = \sigma\!\big(g_{s\rightarrow n}^{(k)}(\bm{U}^{(k)})\big),
	\end{equation}
	where \(g_{\cdot}^{(k)}\) are projections with output channels $4D_{\rm s}$ and \(\sigma(\cdot)\) denotes the activation. The gates modulate cross-branch messages as
	\begin{equation}
		\bm{H}_{n\rightarrow s}^{(k)}=\bm{I}^{{\rm r}k}_{\rm n}\odot \bm{E}_{n\rightarrow s}^{(k)},\qquad
		\bm{H}_{s\rightarrow n}^{(k)}=\bm{I}^{{\rm r}k}_{\rm s}\odot \bm{E}_{s\rightarrow n}^{(k)},
	\end{equation}
	with \(\odot\) the Hadamard product. Residual cross-updates then yield the post-interaction features, which can be expressed as
	\begin{equation}
		\hat{\bm{I}}^{{\rm r}(k+1)}_{\rm s}=\bm{I}^{{\rm r}k}_{\rm s}+\bm{H}_{n\rightarrow s}^{(k)},\qquad
		\hat{\bm{I}}^{{\rm r}(k+1)}_{\rm n}=\bm{I}^{{\rm r}k}_{\rm n}+\bm{H}_{s\rightarrow n}^{(k)},
	\end{equation}
	where $\hat{\bm{I}}^{{\rm r}(k+1)}_{\rm s}, \hat{\bm{I}}^{{\rm r}(k+1)}_{\rm n} \in \mathbb{R}^{B\times T \times (F/4)\times 4D_{\rm s}}$. The interaction module outputs in channel-first can be expressed as
	\begin{align}
		\hat{\bm{I}}^{{\rm r}(k+1)}_{\rm s} \triangleq & \hat{\bm{I}}^{{\rm r}(k+1)}_{\rm s}, \notag \\
		\hat{\bm{I}}^{{\rm r}(k+1)}_{\rm n} \triangleq & \hat{\bm{I}}^{{\rm r}(k+1)}_{\rm n}\in\mathbb{R}^{B\times 4D_{\rm s}\times T\times (F/4)}.
	\end{align}
	Distinct parameter sets \(g_{n\rightarrow s}^{(k)}\) and \(g_{s\rightarrow n}^{(k)}\) are used at every stage. 
	
	Collecting the above, the dual-branch semantic encoder realizes the mapping
	\begin{equation}
		(\bm{s}_{\rm s},\,\bm{s}_{\rm n}) \;=\; f_{\rm dual}\!\big(\bm{I}\;;\,\bm{\epsilon}_{\rm dual}\big),
	\end{equation}
	where $f_{\rm{dual}}(\cdot)$, $\bm{\epsilon}_{\rm{dual}}$ represent the mapping function and parameters of the dual-branch semantic encoder, respectively. For reference, the semantic vectors $\bm{s}_{\rm s}=\hat{\bm{I}}^{{\rm r}4}_{\rm s}$, $\bm{s}_{\rm n}=\hat{\bm{I}}^{{\rm r}4}_{\rm n}$.
	
	The decoupling mechanism drives the two branches to learn complementary, branch-specific embeddings, placing speech and noise in distinct subspaces of the semantic space and thereby achieving feature separation. Concretely, we encourage the learned speech and noise representations to be orthogonal, which can be expressed as
	\begin{align}
		\big\langle \,\widetilde{\bm{s}}_{\rm s}\,,\,\widetilde{\bm{s}}_{\rm n}\,\big\rangle \;\approx\; 0,
	\end{align}
	where $\widetilde{(\cdot)}$ denotes per-channel normalization and $\langle\cdot,\cdot\rangle$ is the Frobenius inner product over the spatiotemporal dimensions. 
	
	\vspace{-2 mm}
	\subsection{Semantic Decoder N} 
	\vspace{-1 mm}
	Semantic decoder N is used during training to provide noise reconstruction supervision and mixture consistency. It is pruned at inference. The core task of the semantic decoder N is to model and analyze the features of environmental noise. It is composed of a transposed convolutional (ConvTrans) layer and conv blocks. This structure can effectively utilize the respective characteristics to achieve the accurate restoration of the environmental noise. It accurately maps the noise statistical features $\bm{s}_{\rm{n}}$ to the estimated noise $\bm{o}_{\rm{n}}$. 
	
	We initialize the inputs as
	\begin{align}
		\bm{h}^{0}_{\rm n} & \triangleq \bm{s}_{\rm n}\in\mathbb{R}^{B\times 4D_{\rm s}\times T\times (F/4)}.
		\notag 
	\end{align}
	\begin{align}
		\bm{d}^{j}_{\rm n} &=
		\begin{cases}
			\bm{I}^{\rm c2'}_{\rm n} \triangleq \bm{I}^{\rm c2}_{\rm n}\in\mathbb{R}^{B\times 2D_{\rm s}\times T\times (F/2)}, & j=1,\\[2pt]
			\bm{I}^{\rm c1'}_{\rm n} \triangleq \bm{I}^{\rm c1}_{\rm n}\in\mathbb{R}^{B\times D_{\rm s}\times T\times F}, & j=2,\\[2pt]
			\bm{I}' \triangleq \pi(\bm{I})\in\mathbb{R}^{B\times C_{\rm in}\times T\times F}, & j=3,
		\end{cases}
	\end{align}
	where $\pi(\cdot)$ denotes a fixed rearrangement that aligns $\bm{I}$ to the decoder’s channel-first layout.
	For the stage-wise update, at each decoding stage $j\in\{1,2,3\}$, the decoder upsamples along the frequency axis and performs gated skip fusion with $\bm{d}^{j}_{\rm n}$. The features can be expressed as
	\begin{align}
		\tilde{\bm{h}}^{j}_{\rm n} &= \mathcal{U}^{j}\!\big(\bm{h}^{(j-1)}_{\rm n}\big),\\
		\bm{g}^{j}_{\rm n} &= \sigma\Big(\mathcal{E}^{\rm n j}_{\rm de}\big([\tilde{\bm{h}}^{j}_{\rm n}\ \|\ \bm{d}^{j}_{\rm n}];\,\boldsymbol{\theta}^{j}_{\rm gate}\big)\Big),\\
		\bm{m}^{j}_{\rm n} &= \bm{d}^{j}_{\rm n}\odot \bm{g}^{j}_{\rm n},\\
		\bm{h}^{j}_{\rm n} &= \mathcal{E}^{\rm n j}_{\rm de}\big([\bm{m}^{j}_{\rm n}\ \|\ \tilde{\bm{h}}^{j}_{\rm n}];\,\boldsymbol{\theta}^{j}_{\rm fuse}\big)\;+\;\tilde{\bm{h}}^{j}_{\rm n},
	\end{align}
	where $\mathcal{U}^{1}\!:\ (F/4)\!\mapsto\!(F/2)$, $\mathcal{U}^{2}\!:\ (F/2)\!\mapsto\!F$, and $\mathcal{U}^{3}$ keeps $F$ unchanged. $[\cdot\|\cdot]$ denotes channel-wise concatenation. $\mathcal{E}^{\rm n j}_{\rm de}(\cdot;\boldsymbol{\theta})$ is the mapping function of a Conv block. Parameter sets $\boldsymbol{\theta}^{j}_{\rm gate}$ and $\boldsymbol{\theta}^{j}_{\rm fuse}$ are used for gate generation and fusion, respectively.
	
	The semantic decoder N outputs the complex ratio mask $\bm{M}_{\rm n}$, which can be expressed as
	\begin{equation}
		\bm{M}_{\rm n} 
		=\bm{h}^{3}_{\rm n} 
		=  f^{\rm n}_{\rm sde}\!\big(\bm{s}_{\rm n},\,\bm{I}^{\rm c2}_{\rm n},\,\bm{I}^{\rm c1}_{\rm n},\,\bm{I}\;;\,\bm{\kappa}_{\rm n}\big),
	\end{equation}
	where $f^{\rm n}_{\rm sde}(\cdot)$ and $\bm{\kappa}_{\rm n}$ denote the mapping function and parameters of the semantic decoder N. $\bm{M}_{\rm n}$ is applied to the mixture spectrum to recover the estimated noise spectrum. This process can be expressed as
	\begin{equation}
		\widehat{\bm{N}} = \bm{M}_{\rm n} \odot \bm{I}. 
	\end{equation}
	The corresponding time domain noise signal is then obtained by the Inverse STFT (ISTFT), which can be expressed as
	\begin{equation}
		\bm{o}_{\rm n} = \mathcal{S}^{-1}(\widehat{\bm{N}}),
	\end{equation}
	where $\mathcal{S}^{-1}(\cdot)$ denotes the mapping function of ISTFT.
	
	\vspace{-2 mm}
	\subsection{Channel Encoder} 
	\vspace{-1 mm}
	
	The semantic vectors $\bm{s}_{\rm s}$ are fed to the channel encoder. The channel encoder greatly enhances the ability of the signal to resist various interferences during the transmission process. According to the characteristics of channels, it converts $\bm{s}_{\rm{s}}$ to a power-constrained transmit signal $t$.  
	
	Before channel encoding, all tensors are first unified to the target frequency resolution $F$, which can be expressed as
	$$
	\bar{\bm{s}}_{\rm s}=\mathrm{pad}_F(\bm{s}_{\rm s}),\quad
	\bar{\bm{I}}^{\rm c2}_{\rm s}=\mathrm{pad}_F(\bm{I}^{\rm c2}_{\rm s}),\quad
	\bar{\bm{I}}^{\rm c1}_{\rm s}=\bm{I}^{\rm c1}_{\rm s},\quad
	\bar{\bm{I}}=\bm{I},
	$$
	where $\mathrm{pad}_F(\cdot)$ denotes zero-padding along the frequency dimension. The resulting channel-first input to the channel encoder is $\bm{h}^{0}_{\rm c}\;\triangleq\; \big[\;\bar{\bm{s}}_{\rm s}\ \|\ \bar{\bm{I}}^{\rm c2}_{\rm s}\ \|\ \bar{\bm{I}}^{\rm c1}_{\rm s}\ \|\ \bar{\bm{I}}\;\big]\ \in\ \mathbb{R}^{B\times(7D_{\rm s}+C_{\rm in})\times T\times F}$.
	Then, the channel encoder takes a channel-first tensor \(\bm{h}^{0}_{\rm c}\) and maps it to the transmit representation through three Conv blocks. The features can be expressed as
	\begin{equation}
		\bm{h}^{j}_{\rm c} \;=\; \mathcal{B}^{\rm cen}_j\!\big(\bm{h}^{(j-1)}_{\rm c};\,\boldsymbol{\zeta}^{j}\big),\quad j=1,2,3,
	\end{equation}
	where \(\mathcal{B}^{\rm cen}_j(\cdot)\) denotes the \(j\)-th Conv block and \(\boldsymbol{\zeta}^{j}\) its parameters.
	The transmit signal $t$ can be expressed as 
	\begin{equation}
		\bm{t} \;=\; \bm{h}^{3}_{\rm c} \;=\; f_{\rm cen}\!\big(\bm{h}^{0}_{\rm c};\,\bm{\zeta}\big),\qquad 
		\bm{\zeta}\triangleq\{\boldsymbol{\zeta}^{j}\}_{j=1}^3,
	\end{equation}
	where $f_{\rm{cen}}(\cdot)$, $\bm{\zeta}$ represent the mapping function and parameters of the channel encoder. Power normalization enforces the transmit constraint, which can be expressed as
	\begin{equation}
		\bm{t}\ \leftarrow\ \sqrt{P}\;\frac{\bm{t}}{\sqrt{\mathbb{E}\!\left[\|\bm{t}\|_2^2\right]}}\quad\text{so that}\quad \mathbb{E}\!\left[\|\bm{t}\|_2^2\right]\le P,
	\end{equation}
	where $P$ denotes the target average transmit power budget, the expectation $\mathbb{E}[\cdot]$ is taken over the training distribution, and $\|\cdot\|_2$ is the Euclidean norm.
	
	The transmit signal $\bm{t}$ is vectorized and sent over the channel. Channel effects are a key external factor determining E2E quality. We model the link as a flat-fading channel acting on the vectorized baseband signal. The received signal $r$ can be expressed as
	\begin{equation}
		\bm{r}\;=\;h\,\bm{t}+\bm{n},\qquad
		h\sim\mathcal{CN}(0,1),\ \ \bm{n}\sim\mathcal{CN}(\bm{0},\,\sigma^2\bm{I}),
	\end{equation}
	where $h$ is the channel fading coefficient, which reflects the attenuation effect of the channel on the signal strength. $\bm{n}$ is the additive noise, which includes various random noise interferences. These factors work together, causing a difference between the received signal $r$ and the transmission signal $t$.

	\subsection{Channel Decoder}  
	\vspace{-1 mm}

	The channel decoder adjusts the decoding strategy according to the characteristics of the received signal. It effectively suppresses channel noise interference to semantic features. Moreover, it maps $\bm{r}$ to the target speech semantic representation $\bm{x}_{\rm s}$, which is as close as possible to the original target speech semantic vectors and supports speech reconstruction. 
	
	It mirrors the channel encoder with three Conv blocks, thereby preserving \(T\) and \(F\) while progressively refining the received representation. Let the received code be $\bm{h}^{0}_{\rm d}\triangleq \bm{r}$. The updates are
	\begin{equation}
		\bm{h}^{j}_{\rm d} \;=\; \mathcal{B}^{\rm cde}_j\!\big(\bm{h}^{(j-1)}_{\rm d};\,\boldsymbol{\eta}^{j}\big),\quad j=1,2,3,
	\end{equation}
	where \(\mathcal{B}^{\rm cde}_j(\cdot)\) denotes the \(j\)-th decoder Conv block with parameters \(\boldsymbol{\eta}^{j}\). The decoded semantic features are
	\begin{equation}
		\bm{x}_{\rm s} \;=\; \bm{h}^{3}_{\rm d} \;=\; f_{\rm cde}\!\big(\bm{r};\,\bm{\eta}\big),\qquad 
		\bm{\eta}\triangleq\{\boldsymbol{\eta}^{j}\}_{j=1}^3.
	\end{equation}
	where $f_{\rm{cde}}(\cdot)$, $\bm{\eta}$ represent the mapping function and parameters of the channel decoder. 
	
	\vspace{-2 mm}
	\subsection{Semantic Decoder S}     
	\vspace{-1 mm}
	The main responsibility of the semantic decoder S is to reconstruct the semantic features $\bm{x}_{\rm{s}}$ with high fidelity to restore a high-quality target speech. The semantic decoder S adopts the same structure as the semantic decoder N and fully excavates the key information in $\bm{x}_{\rm{s}}$. It converts the semantic features $\bm{x}_{\rm{s}}$ into the reconstructed speech $\bm{o}_{\rm{s}}$ with high semantic integrity and naturalness.
	
	We take the output of the channel decoder and split it along channels as $\bm{x}_{\rm s}=\big[\;\bar{\bm{s'}}_{\rm s}\ \|\ \bar{\bm{I}}_{\rm s}^{\rm c2'}\ \|\ \bar{\bm{I}}_{\rm s}^{\rm c1'}\ \|\ \bar{\bm{I}'}\;\big]$, where $\bar{\cdot}$ indicates the segments extracted from $\bm{x}_{\rm s}$. We then undo the earlier zero-padding by cropping along the frequency axis to recover the original resolutions, which can be expressed as
	\begin{align}
		\bm{h}^{0}_{\rm s} & \triangleq \bar{\bm{s'}}_{\rm s}\in\mathbb{R}^{B\times 4D_{\rm s}\times T\times (F/4)}.
		\notag 
	\end{align}
	\begin{align}
		\bm{d}^{j}_{\rm s} &=
		\begin{cases}
			\bm{I}^{\rm c2''}_{\rm s} \triangleq \bar{\bm{I}}^{\rm c2'}_{\rm s}\in\mathbb{R}^{B\times 2D_{\rm s}\times T\times (F/2)}, & j=1,\\[2pt]
			\bm{I}^{\rm c1''}_{\rm s} \triangleq \bar{\bm{I}}^{\rm c1'}_{\rm s}\in\mathbb{R}^{B\times D_{\rm s}\times T\times F}, & j=2,\\[2pt]
			\bm{I}'' \triangleq \pi(\bar{\bm{I'}})\in\mathbb{R}^{B\times C_{\rm in}\times T\times F}, & j=3.
		\end{cases}
	\end{align}
	Then, semantic decoder S follows the same three stage updates as semantic decoder N, replacing the subscript ${\rm n}$ to ${\rm s}$, using the same upsamplers $\mathcal{U}^{j}$ and gated skip-fusion blocks $\mathcal{E}^{\rm s\,j}_{\rm de}(\cdot)$ for $j=1,2,3$. The semantic decoder S outputs the complex ratio mask $\bm{M}_{\rm s}$, which can be expressed as
	\begin{equation}
		\bm{M}_{\rm s} = \bm{h}^{3}_{\rm s} 
		= f^{\rm s}_{\rm sde}\!\big(\bar{\bm{s'}}_{\rm s},\,\bar{\bm{I}}^{\rm c2'}_{\rm s},\,\bar{\bm{I}}^{\rm c1'}_{\rm s},\,\bar{\bm{I}'}\;;\,\bm{\kappa}_{\rm s}\big),
	\end{equation}
	where $f^{\rm s}_{\rm sde}(\cdot)$ and $\bm{\kappa}_{\rm s}$ represent the mapping function and parameters of the semantic decoder S, respectively. $\bm{M}_{\rm s}$ is applied to the mixture spectrum to obtain the estimated target speech spectrum
	\begin{equation}
		\widehat{\bm{S}} = \bm{M}_{\rm s} \odot \bar{\bm{I}'}.
	\end{equation}
	The reconstructed target speech waveform is then obtained by the ISTFT, which can be expressed as
	\begin{equation}
		\bm{o}_{\rm s} = \mathcal{S}^{-1}(\widehat{\bm{S}}).
	\end{equation}
	
	\vspace{-2 mm}
	\subsection{Loss Function}
	\vspace{-1 mm}

	Guided by semantic information theory, the proposed S-DSA aims to preserve the semantics of the target speech while suppressing noise leakage in the learned representations. The ideal optimization objective is
	\begin{align}
		\label{eq:mi_objective}
		\max\ & I(\bm{s}_{\rm s};\bm{i}_{\rm s}) - \beta\, I(\bm{s}_{\rm s};\bm{i}_{\rm n}) \\[2pt]
		\text{s.t.} \quad & \mathbb{E}\!\big[\|\bm{t}\|_2^2\big]\le P, \notag \\
		& I(\bm{s}_{\rm s}; \bm{s}_{\rm n}) \approx 0, \notag
	\end{align}
	where $I(\cdot\,;\cdot)$ denotes mutual information, $\bm{t}$ is the transmit signal with power budget $P$, and $\beta\!\geq\!0$ balances semantic preservation and noise suppression.  
	
	Since direct optimization of~\eqref{eq:mi_objective} is intractable, we approximate it with a weighted multi-term loss that enforces semantic fidelity, noise decoupling, and perceptual quality. The overall training loss is
	\begin{equation}
		\label{loss}
		L_{\rm total} = \alpha_1 L_{\rm s} + \alpha_2 L_{\rm n} + \alpha_3 L_{\rm mix} + \alpha_4 L_{\rm time} + \alpha_5 L_{\rm mr},
	\end{equation}
	where $\alpha_1,\dots,\alpha_5$ are non-negative coefficients. The five loss components promote accurate semantic reconstruction, effective noise suppression, mixture consistency, waveform fidelity, and perceptual quality.
	
	From a representation perspective, $L_{\rm s}$, $L_{\rm n}$, and $L_{\rm mix}$ operate on the frequency domain to supervise target speech reconstruction, noise estimation, and mixture consistency on spectrograms. Specifically, the speech reconstruction loss $L_{\rm s}$ minimizes the discrepancy between the estimated and clean speech spectra
	\begin{equation}
		L_{\rm s} = \mathbb{E}\!\left[\|\widehat{\bm{S}}-\bm{S}\|_2^2\right].
	\end{equation} The noise estimation loss $L_{\rm n}$ supervises the noise branch to recover the noise spectrum accurately
	\begin{equation}
		L_{\rm n} = \mathbb{E}\!\left[\|\widehat{\bm{N}}-\bm{N}\|_2^2\right].
	\end{equation} 
	The mixture consistency loss $L_{\rm mix}$ encourages the reconstructed speech and noise to sum to the observed mixture
	\begin{equation}
		L_{\rm mix} = \mathbb{E}\!\left[\|\widehat{\bm{S}}+\widehat{\bm{N}}-\bm{Y}\|_2^2\right].
	\end{equation}
	
	In contrast, $L_{\rm time}$ and $L_{\rm mr}$ are waveform oriented criteria. The former penalizes sample wise errors in the time domain, while the latter matches multi resolution spectral envelopes computed from the waveform.
	The time domain loss $L_{\rm time}$ constrains waveform-level reconstruction quality
	\begin{equation}
		L_{\rm time} = \mathbb{E}\!\left[\|\bm{o}_{\rm s}-\bm{i}_{\rm s}\|_1\right].
	\end{equation} 
	The multi-resolution STFT loss $L_{\rm mr}$ evaluates the similarity between $\bm{o}_{\rm s}$ and $\bm{i}_{\rm s}$ under $M$ different Fast Fourier Transform (FFT) configurations. For $m$-th resolution
	\begin{align}
		\widehat{\bm{S}}^{(m)} &= \mathrm{STFT}_m(\bm{o}_{\rm s}), \\
		\bm{S}^{(m)} &= \mathrm{STFT}_m(\bm{i}_{\rm s}),
	\end{align}
	where $\mathrm{STFT}_m(\cdot)$ denotes the STFT computed with $m$-th configuration. The spectral convergence $\mathrm{SC}_m$ and log-magnitude $\mathrm{Lmag}_m$ are defined as
	\begin{align}
		\mathrm{SC}_m &= 
		\frac{\|\,|\widehat{\bm{S}}^{(m)}|-|\bm{S}^{(m)}|\,\|_F}
		{\|\,|\bm{S}^{(m)}|\,\|_F}, \\[3pt]
		\mathrm{Lmag}_m &= 
		\|\log(|\widehat{\bm{S}}^{(m)}|+\epsilon)-\log(|\bm{S}^{(m)}|+\epsilon)\|_1,
	\end{align}
	where $|\cdot|$ is the magnitude spectrogram, $\|\cdot\|_F$ the Frobenius norm, $\|\cdot\|_1$ the elementwise $\ell_1$ norm, and $\epsilon$ a small constant for numerical stability. The overall multi-resolution STFT loss is the average across resolutions
	\begin{equation}
		L_{\rm mr} = \frac{1}{M}\sum_{m=1}^{M}(\mathrm{SC}_m+\mathrm{Lmag}_m).
	\end{equation}
	
	\vspace{-1 mm}
	\section{Algorithm Design}
	\vspace{-1 mm}
	
	\begin{algorithm}[b!]
		\caption{Training Algorithm}
		\label{alg:train}
		\begin{algorithmic}[1]
			\State \textbf{Input:} Mini-batches of $(\bm{i}_{\rm s}, \bm{i}_{\rm n})$; optimizer $\mathcal{O}$; channel SNR sampler.
			\State \textbf{Output:} Trained parameters $\Theta^\star$.
			
			\State Initialize network parameters $\Theta$.
			\While{not converged}
			
			\State $\bm{i}\gets \bm{i}_{\rm s}+\bm{i}_{\rm n}$ \Comment{mixture construction}
			\State $\bm{Y}\gets \mathcal{S}(\bm{i}),\ \bm{S}\gets \mathcal{S}(\bm{i}_{\rm s}),\ \bm{N}\gets \mathcal{S}(\bm{i}_{\rm n})$ \Comment{STFT}
			\State $\bm{I}\gets [\,\Re(\bm{Y})\ \|\ \Im(\bm{Y})\,]$
			
			\Statex \textit{\qquad // Transmitter}
			\State $(\bm{s}_{\rm s},\bm{s}_{\rm n}) \gets f_{\rm dual}(\bm{I};\,\bm{\epsilon}_{\rm dual})$
			
			\State $\bm{M}_{\rm n} \gets f^{\rm n}_{\rm sde}(\bm{s}_{\rm n},\,\bm{I}^{\rm c2}_{\rm n},\,\bm{I}^{\rm c1}_{\rm n},\,\bm{I};\,\bm{\kappa}_{\rm n})$ \Comment{noise mask}
			\State $\widehat{\bm{N}}\gets \bm{M}_{\rm n}\odot \bm{I}$,
			\State $\bm{o}_{\rm n}\gets \mathcal{S}^{-1}(\widehat{\bm{N}})$ \Comment{estimated noise}
			\State $\bm{t}\gets f_{\rm cen}(\bm{s}_{\rm s};\,\bm{\zeta})$; 
			\State $\bm{t}\leftarrow \mathrm{PowerNorm}(\bm{t},P)$

			\Statex \textit{\qquad // Channel}
			\State Sample channel condition $h$ and noise $\bm{n}$
			\State $\bm{r}\gets h\,\bm{t}+\bm{n}$ 
			
			\Statex \textit{\qquad // Receiver}
			\State $\bm{x}_{\rm s}\gets f_{\rm cde}(\bm{r};\,\bm{\eta})$
			\State $\bm{M}_{\rm s}\gets f^{\rm s}_{\rm sde}(\bar{\bm{s'}}_{\rm s},\,\bar{\bm{I}}^{\rm c2'}_{\rm s},\,\bar{\bm{I}}^{\rm c1'}_{\rm s},\,\bar{\bm{I}'};\,\bm{\kappa}_{\rm s})$ \Comment{speech mask}
			\State $\widehat{\bm{S}}\gets \bm{M}_{\rm s}\odot \bar{\bm{I}'}$,
			\State $\bm{o}_{\rm s}\gets \mathcal{S}^{-1}(\widehat{\bm{S}})$ \Comment{estimated speech}

			\Statex \textit{\qquad // Loss and update}
			\State Compute the total loss \ref{loss}.
			\State Update $\Theta \gets \mathcal{O}\big(\Theta, \nabla_\Theta L_{\rm total}\big)$.
			\State Apply adaptive learning rate scheduling and check early stopping.
			\EndWhile
			\State \textbf{return} $\Theta^\star$
		\end{algorithmic}
	\end{algorithm}
	
	This section presents the training and testing algorithms. The training algorithm jointly optimizes feature decoupling and semantic reconstruction. The testing algorithm retains the dual-branch encoder for robust feature decoupling, while disabling the noise decoder for lightweight inference.
	
	\vspace{-4 mm}
	\subsection{Training Algorithm}
	\vspace{-1 mm}

	The training algorithm follows the loss in Eq. \ref{loss} and updates all modules jointly as Algorithm \ref{alg:train}. Specifically, the input is a batch of $\bm{i}_{\rm s}$ and $\bm{i}_{\rm n}$. The output is the trained parameter set $\Theta^\star=\{\bm{\epsilon}_{\rm dual},\ \bm{\kappa}_{\rm s},\ \bm{\kappa}_{\rm n},\ \bm{\zeta},\ \bm{\eta}\}$.

	For each batch, the transmitter takes $\bm{i}_{\rm{s}}$ and $\bm{i}_{\rm{n}}$ as inputs. The dual-branch semantic encoder extracts the features of speech and noise in parallel to perform dual-path feature decoupling, obtaining features $\bm{s}_{\rm{s}}, \bm{s}_{\rm{n}}$. $\bm{s}_{\rm{n}}$ is used for noise estimation through the semantic decoder N. $\bm{s}_{\rm{s}}$ is encoded by the channel encoder to obtain $t$ and transmitted through the channel. The receiver receives the signal $r$. The channel decoder decodes it to recover the semantic features $\bm{x}_{\rm{s}}$ of the target speech. The semantic decoder S reconstructs the speech $\bm{o}_{\rm{s}}$ and outputs it. 
	
	The total loss $L_{\rm total}$ is used to update the parameters $\Theta$ by backpropagation over all batches. To enhance training stability and generalization, channel SNRs are sampled using a dual-anchor strategy that covers both low and high channel conditions. An adaptive learning rate scheduler accelerates convergence in the early stage and ensures fine-grained optimization in later epochs, while early stopping prevents overfitting when validation loss no longer improves.

	\begin{algorithm}[b!]
		\caption{Testing Algorithm}
		\label{alg:test}
		\begin{algorithmic}[1]
			\State \textbf{Input:} Mixed speech $\bm{i}'$; trained parameters $\Theta^\star$; channel SNRs.
			\State \textbf{Output:} Reconstructed target speech $\bm{o}'_{\rm s}$.
			\For{each channel SNR}
			
			\State $\bm{Y}'\gets \mathcal{S}(\bm{i}')$;
			\State $\bm{I}'\gets [\,\Re(\bm{Y}')\ \|\ \Im(\bm{Y}')\,]$
			
			\Statex \textit{\qquad // Transmitter}
			
			\State $(\bm{s}'_{\rm s},\bm{s}'_{\rm n}) \gets f_{\rm dual}(\bm{I}';\,\bm{\epsilon}_{\rm dual})$
			\State $\bm{t}'\gets f_{\rm cen}(\bm{s}'_{\rm s};\,\bm{\zeta})$;
			\State $\bm{t}'\leftarrow \mathrm{PowerNorm}(\bm{t}',P)$

			\Statex \textit{\qquad // Channel}
			\State Sample $h',\bm{n}'$ per SNR;
			\State $\bm{r}'\gets h'\,\bm{t}'+\bm{n}'$
			
			\Statex \textit{\qquad // Receiver}
			\State $\bm{x}'_{\rm s}\gets f_{\rm cde}(\bm{r}';\,\bm{\eta})$
			
			\State $\bm{M}'_{\rm s}\gets f^{\rm s}_{\rm sde}(\bar{\bm{s''}}_{\rm s},\,\bar{\bm{I}'}^{\rm c2'}_{\rm s},\,\bar{\bm{I}'}^{\rm c1'}_{\rm s},\,\bar{\bm{I}''};\,\bm{\kappa}_{\rm s})$

			\State $\widehat{\bm{S}}'\gets \bm{M}'_{\rm s}\odot \bar{\bm{I}''}$,

			\State $\bm{o}'_{\rm s}\gets \mathcal{S}^{-1}(\widehat{\bm{S}}')$
			
			\State \textbf{Output} $\bm{o}'_{\rm s}$
			\EndFor
		\end{algorithmic}
	\end{algorithm}
	
	\vspace{-2 mm}
	\subsection{Testing Algorithm}
	\vspace{-1 mm}

	The testing algorithm is shown in Algorithm \ref{alg:test}. The semantic decoder N is disabled to reduce latency and memory, while the dual-branch semantic encoder with stage-wise interaction remains active to retain semantic noise decoupling. The input is the mixed speech $\bm{i}'$. The proposed S-DSA performs E2E transmission and reconstruction of the target speech only.
	
	For each channel SNR, the input is the mixed speech $\bm{i}$ and the trained parameters $\Theta$, and the output is the reconstructed target speech $\bm{o}'_s$. The transmitter takes the target speech $\bm{i}'_s$ and the environmental noise $\bm{i}'_n$ as inputs. The dual-branch semantic encoder extracts the features of speech and noise in parallel, obtaining the semantic features $\bm{s}'_s, \bm{s}'_n$. $\bm{s}'_s$ is encoded by the channel encoder to obtain $t'$ and transmitted through the channel. The receiver receives the signal $r'$. The channel decoder decodes the received signal to recover the semantic features $\bm{x}'_s$ of the target speech. The semantic decoder S reconstructs the speech $\bm{o}'_s$ and outputs it.
	
	Compared with training, skipping noise decoding achieves lightweight inference and reduces computational overhead, while the encoder-side feature decoupling ensures that robustness against environmental noise is preserved. 
	
	\vspace{-1 mm}
	\section{Simulation Experiments}
	\vspace{-1 mm}
	
	In this section, we describe the simulation setup, evaluation metrics, and benchmarks. We then compare the proposed S-DSA with benchmarks on various channels, considering different environmental noise.
	
	\subsection{Simulation Setup}
	\vspace{-1 mm}
	
	The simulation parameter settings of the proposed S-DSA are shown in Tab. \ref{settings}.  The implementation uses PyTorch and the Adam optimizer. Training is conducted for up to 120 epochs with early stopping. The learning rate is initialized to $10^{-3}$ and is reduced by a validation-driven scheduler to a minimum of $10^{-4}$.
	Audio is sampled at 16 kHz. The STFT uses a 512-point FFT and a hop of 256 samples with a Hann window.
	
	\begin{table}[h]
		\centering
		\caption{Simulation parameter settings of the proposed S-DSA.}
		\begin{tabularx}{0.46\textwidth}{p{4cm}p{6cm}}
			\hline
			\textbf{Parameter} & \textbf{Setting} \\
			\hline
			Implementation framework & PyTorch \\
			Optimizer & Adam \\
			Max training epochs & 120 \\
			
			Learning rate & Initial $10^{-3}$, minimum $10^{-4}$ \\
			
			Batch size & 32 \\
			Validation split & 20\% \\
			Audio sampling rate (kHz) & 16  \\
			
			STFT FFT size & 512 \\
			STFT hop size & 256 \\
			
			Environmental noise & AWGN \\
			Channel model & AWGN, Rician, Rayleigh \\
			Training $\mathrm{SNR}_{\text{in}}$ (dB) & 10  \\
			Training $\mathrm{SNR}_{\text{ch}}$ (dB) & 3  and 15 \\
			Testing $\mathrm{SNR}_{\text{in}}$ or $\mathrm{SNR}_{\text{ch}}$ (dB) & [-10, -5, 0, 5, 10, 15, 20] \\
			\hline
		\end{tabularx}
		\label{settings}
	\end{table}
	
	We explicitly distinguish between two SNR definitions. The input mixing SNR, $\mathrm{SNR}_{\text{in}}$, controls the energy ratio between $\bm{i}_{\rm s}$ and $\bm{i}_{\rm n}$ when forming the mixture $\bm{i}=\bm{i}_{\rm s}+\bm{i}_{\rm n}$. The channel SNR, $\mathrm{SNR}_{\text{ch}}$, characterizes the link quality in $\bm{r}=h\bm{t}+\bm{n}$. $\mathrm{SNR}_{\text{in}}$ is fixed at 10 dB during training and validation. $\mathrm{SNR}_{\text{ch}}$ is sampled from two anchors near 3 dB and 15 dB to cover low and high link qualities. Environmental noise is generated as AWGN in the time domain to match the target SNR. The physical channel is configured as AWGN, Rician fading, or Rayleigh fading. During evaluation, both the channel SNR and the input mixing SNR are varied across a wide range to test system robustness.
	
	
	To comprehensively evaluate the proposed S-DSA, we adopt complementary metrics that characterize both waveform fidelity and perceptual quality. For waveform fidelity, we report the mean squared error (MSE) between the enhanced and clean speech, as well as the scale-invariant signal-to-distortion ratio (SI-SDR) \cite{PESQ}. For perceptual quality, we employ the Perceptual Evaluation of Speech Quality (PESQ) \cite{PESQNEW} and the Short-Time Objective Intelligibility (STOI) \cite{STOI}, which are widely used to assess speech quality and intelligibility.
	
	
	We compare the following benchmarks, all trained under the same data protocol and optimization settings to reconstruct the clean target speech.
	
	(1) CNN-AE: A convolutional autoencoder communication system. It directly compresses the input mixture and reconstructs the speech signal at the receiver, without an explicit noise branch.
	
	(2) DeepSC-S: An attention enhanced semantic communication system with squeeze and excitation modules. It reconstructs speech features from the mixture \cite{DeepSC-S}.
	
	(3) S-DSA@Oracle: The proposed S-DSA is evaluated under an ideal noiseless, nonfading link with unit channel gain and zero noise, such that $\bm{r}$ is identical to $\bm{t}$.
	
	\subsection{Simulation Results}
	
	\subsubsection{Speech Enhancement Visualization}
	
\begin{figure}[b!]
		\centering
		\includegraphics[width=1\linewidth]{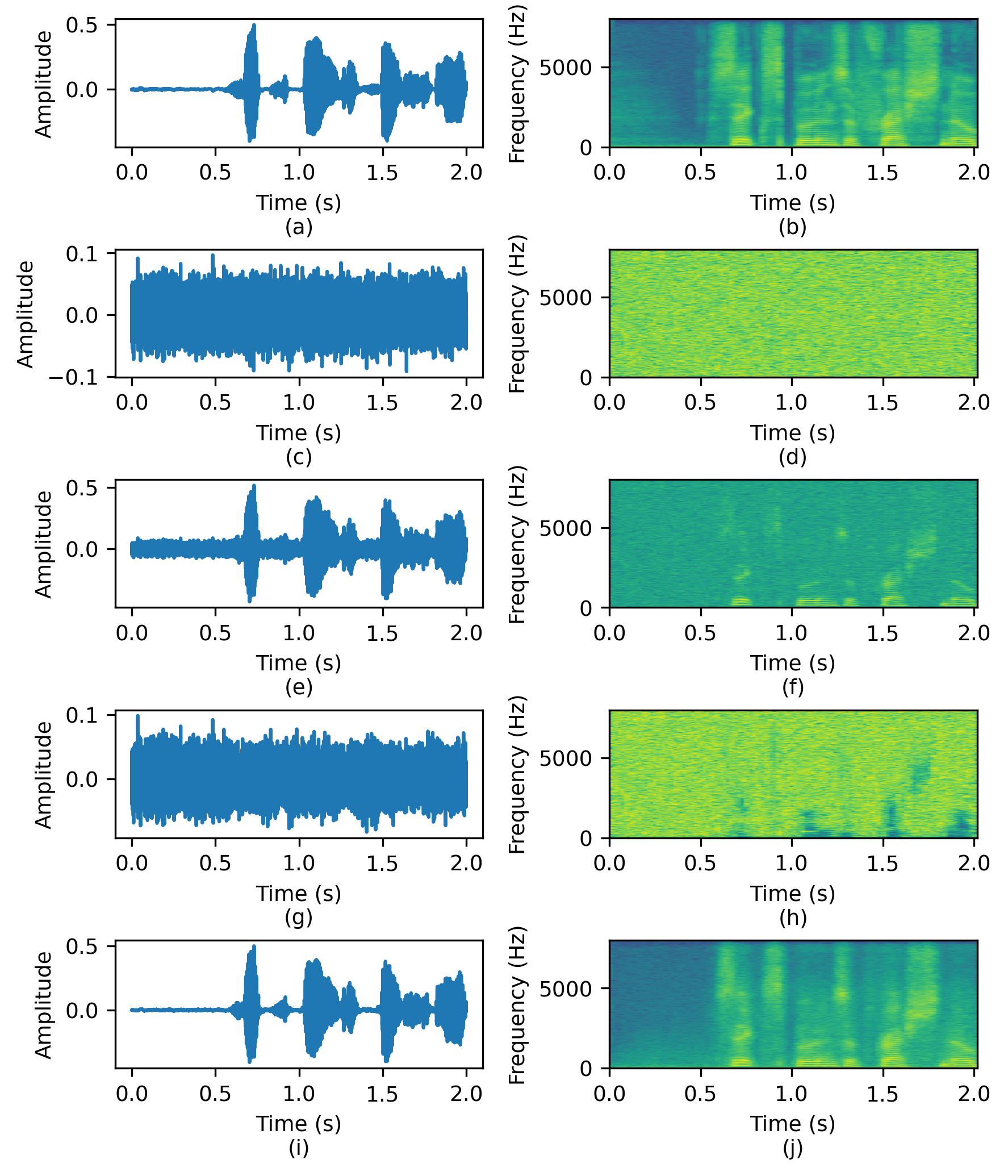}
		\caption{Waveform and spectrogram comparisons for speech enhancement.}
		\label{visualization}
	\end{figure}

	Fig. \ref{visualization} illustrates the waveforms and spectrograms produced by the proposed S-DSA at different stages of speech processing. From top to bottom, the panels show: (a, b) clean speech, (c, d) additive environmental noise, (e, f) mixed input, (g, h) estimated noise, and (i, j) enhanced target speech.
	
	In the time domain, clean speech exhibits clear periodic patterns with silent gaps, whereas environmental noise shows irregular low-amplitude fluctuations. Once corrupted by noise, the mixed input loses regularity and displays severe distortion. The proposed S-DSA effectively separates the noise component, reducing its residual amplitude to a minimal level, while the enhanced speech waveform recovers both the amplitude range and temporal periodicity of the clean reference.
	
	%
	%
	%
	
	\begin{figure*}[t]
		\centering
		\subfloat[AWGN\label{AWGN_snr_channel}]%
		{\includegraphics[width=1\textwidth]{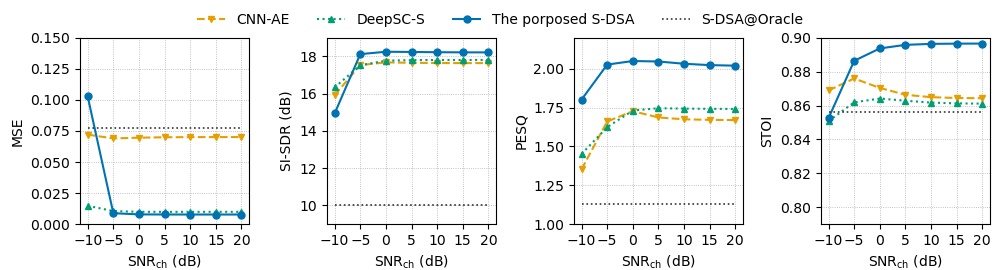}}\par
		\subfloat[Rician\label{Rician_snr_channel}]%
		{\includegraphics[width=1\textwidth]{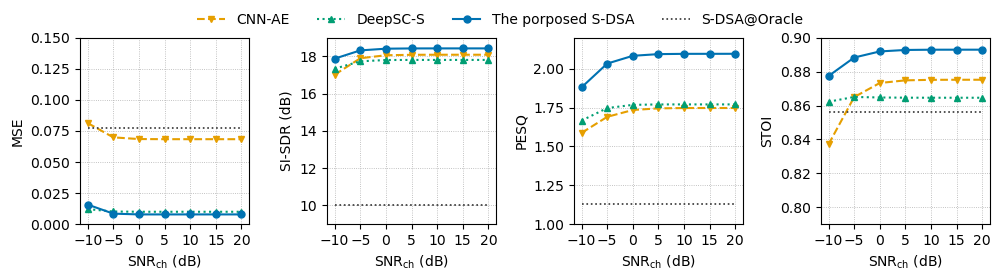}}\par
		\subfloat[Rayleigh\label{Rayleigh_snr_channel}]%
		{\includegraphics[width=1\textwidth]{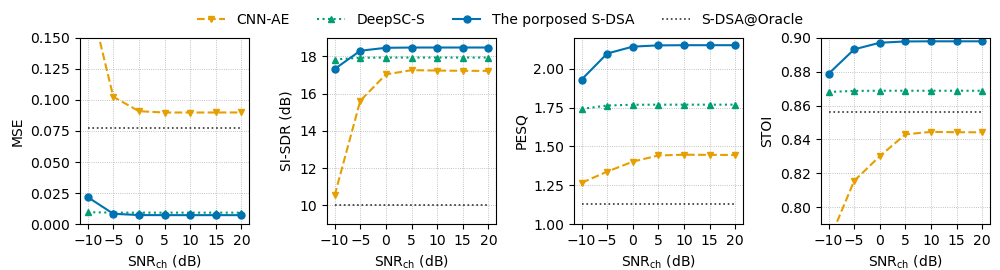}}
		\caption{Performance metrics versus channel SNR across different channels.}
		\label{snr_channel}
	\end{figure*}

	\begin{figure*}[t]
		\centering
		\includegraphics[width=1\linewidth]{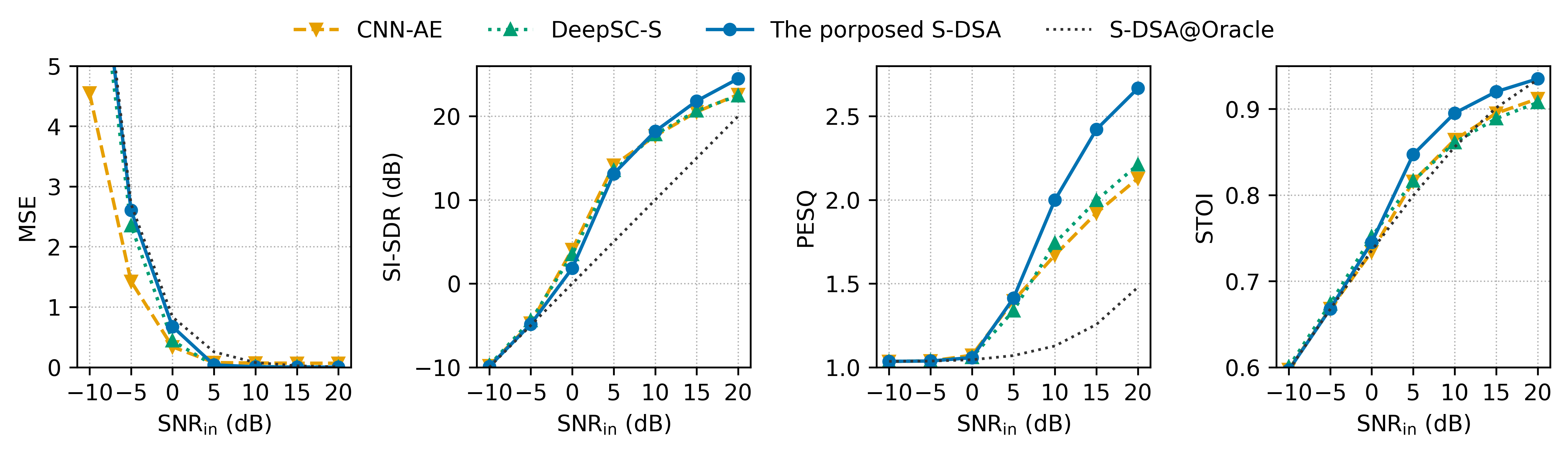}
		\caption{Performance metrics versus environmental SNR on the AWGN channel.}
		\label{snr_input}
	\end{figure*}
	
	The spectrogram analysis further highlights the advantages of the proposed S-DSA. Clean speech energy is mainly concentrated in the 500–3000~Hz region with distinct formant structures, while environmental noise spreads energy uniformly across the spectrum. As a result, the mixed input suffers from broadened bandwidth and blurred formant peaks. In contrast, the proposed S-DSA output successfully restores sharp formant structures and a concentrated energy distribution in the speech-dominant frequency bands, while relegating noise energy to the separated noise output.
	
	Overall, these visual results confirm that the proposed S-DSA achieves strong noise suppression and accurate speech reconstruction in both time and frequency domains, significantly outperforming the mixed input. Furthermore, the results demonstrate the effective decoupling of speech and noise.
	
	\subsubsection{Adaptability Across Transmission Channels}
	
	Fig.~\ref{snr_channel} presents the performance of the proposed S-DSA, CNN-AE, and DeepSC-S across AWGN, Rician, and Rayleigh channels. The environmental noise of the input signal is fixed at 10 dB. This experiment is designed to evaluate how systems adapt to diverse channel conditions.
	
	Across all channels, the overall trend is consistent. As the channel SNR increases, the MSE decreases steadily, while SI-SDR, PESQ, and STOI scores improve. Although every system benefits from better channel conditions, the proposed S-DSA provides stronger perceptual quality and intelligibility, especially under low-SNR transmission conditions.
	
	On the AWGN channel, the proposed S-DSA shows a faster reduction in MSE compared with CNN-AE and maintains higher SI-SDR values across most SNR levels. At 0 dB, the PESQ score of S-DSA reaches 2.05, which is clearly higher than the 1.73 obtained by both CNN-AE and DeepSC-S. At the same channel SNR, the STOI score of S-DSA rises to 0.89, demonstrating that it can effectively preserve intelligibility under noisy conditions.
	
	On the Rician channel, where a partial line-of-sight component is present, the proposed S-DSA maintains competitive fidelity while surpassing benchmarks in perceptual metrics. When the channel SNR is 0 dB, the PESQ score of S-DSA exceeds 2.08 and the STOI score approaches 0.89, both of which are higher than benchmarks. These results show that the proposed S-DSA benefits from its decoupling mechanism, which suppresses multipath interference more effectively.
	
	On the Rayleigh channel, which represents severe fading without line-of-sight, the superiority of the proposed S-DSA becomes more pronounced. At 0 dB, the PESQ score of S-DSA rises to 2.14, while DeepSC-S achieves 1.77 and CNN-AE only 1.40. At the same time, the STOI score of S-DSA reaches 0.90, surpassing benchmarks. This demonstrates that the proposed S-DSA is capable of maintaining high perceptual quality and intelligibility even in the Rayleigh fading channel.
	
	Overall, the proposed S-DSA demonstrates strong adaptability across different transmission channels by explicitly decoupling speech content from disturbances.  By limiting noise leakage in the latent representation, the proposed S-DSA curbs error propagation to the receiver. Training under mixed channel SNR conditions further improves robustness across scenarios.  The gains are most pronounced at low channel SNRs and under severe fading, where multipath-induced amplitude and phase fluctuations would otherwise erode perceptual cues.  In Rician channels with a partial line-of-sight component, the proposed S-DSA also exploits the more stable path to better preserve intelligibility.

	\subsubsection{Robustness to Environmental Noise}

	Fig. \ref{snr_input} shows the performance comparison of the communication systems on the AWGN channel with a fixed channel SNR of 20 dB, while the input mixing SNR varies from -10 dB to 20 dB. This experiment evaluates how systems adapt to varying environmental noise intensity.
	
	At an extremely low input mixing SNR of –10 dB, which rarely occurs in practical scenarios, all systems suffer from severe degradation. Since such extreme cases are rare, the following analysis focuses on 0 dB and above, which better represent practical transmission environments.
	
	For waveform fidelity, the MSE values of all systems drop rapidly once the input mixing SNR exceeds 0 dB. At 5 dB, the proposed S-DSA achieves an MSE of 0.039, which is much lower than CNN-AE at 0.077 and close to DeepSC-S at 0.029. When the input mixing SNR reaches 20 dB, the proposed S-DSA further reduces its MSE to 0.0024, outperforming CNN-AE at 0.066 and DeepSC-S at 0.0046. This demonstrates the efficiency of the proposed S-DSA in reconstructing accurate speech waveforms when noise conditions improve.
	
	For signal recovery, SI-SDR values increase steadily across all systems. At 10 dB, the proposed S-DSA reaches 18.20, higher than CNN-AE at 17.64. At 20 dB, its SI-SDR further climbs to 24.48, establishing a clear lead over CNN-AE at 22.58 and DeepSC-S at 22.46. These results highlight that the proposed S-DSA restores speech signals more effectively as the environmental noise weakens.
	
	For perceptual quality, the improvements are even more pronounced. At 5 dB, the proposed S-DSA achieves a PESQ of 1.41, which is higher than CNN-AE at 1.40 and DeepSC-S at 1.34. At 10 dB, the proposed S-DSA rises to 2.00, while CNN-AE and DeepSC-S remain at 1.67 and 1.74 respectively. At 20 dB, the proposed S-DSA reaches 2.67, outperforming CNN-AE at 2.13 and DeepSC-S at 2.21. This demonstrates that the proposed S-DSA consistently provides the best perceived quality across the entire range of noise conditions.
	
	For intelligibility, the STOI results confirm the same advantage. At 5 dB, the proposed S-DSA achieves 0.847, clearly higher than CNN-AE and DeepSC-S, both around 0.816. At 10 dB, it rises to 0.895, again above the baselines. At 20 dB, the proposed S-DSA reaches 0.935, while CNN-AE and DeepSC-S remain below 0.912. This shows that the proposed S-DSA ensures intelligibility gains across both medium and high input mixing SNR conditions.
	
	The proposed S-DSA exhibits strong robustness against varying environmental noise intensities. It remains competitive even in extremely harsh conditions. Notably, within the practically relevant SNR range of 0–20 dB, it consistently outperforms the benchmarks across reconstruction fidelity, recovery accuracy, perceptual quality, and intelligibility. This advantage stems from semantic-noise feature decoupling, which suppresses noise while retaining essential speech features. Consequently, the proposed S-DSA goes beyond waveform reconstruction and instead optimizes perceptual speech enhancement for a better overall listening experience.
	
	\subsubsection{Training Convergence}
	
	\begin{figure}[t]
		\centering
		\includegraphics[width=1\linewidth]{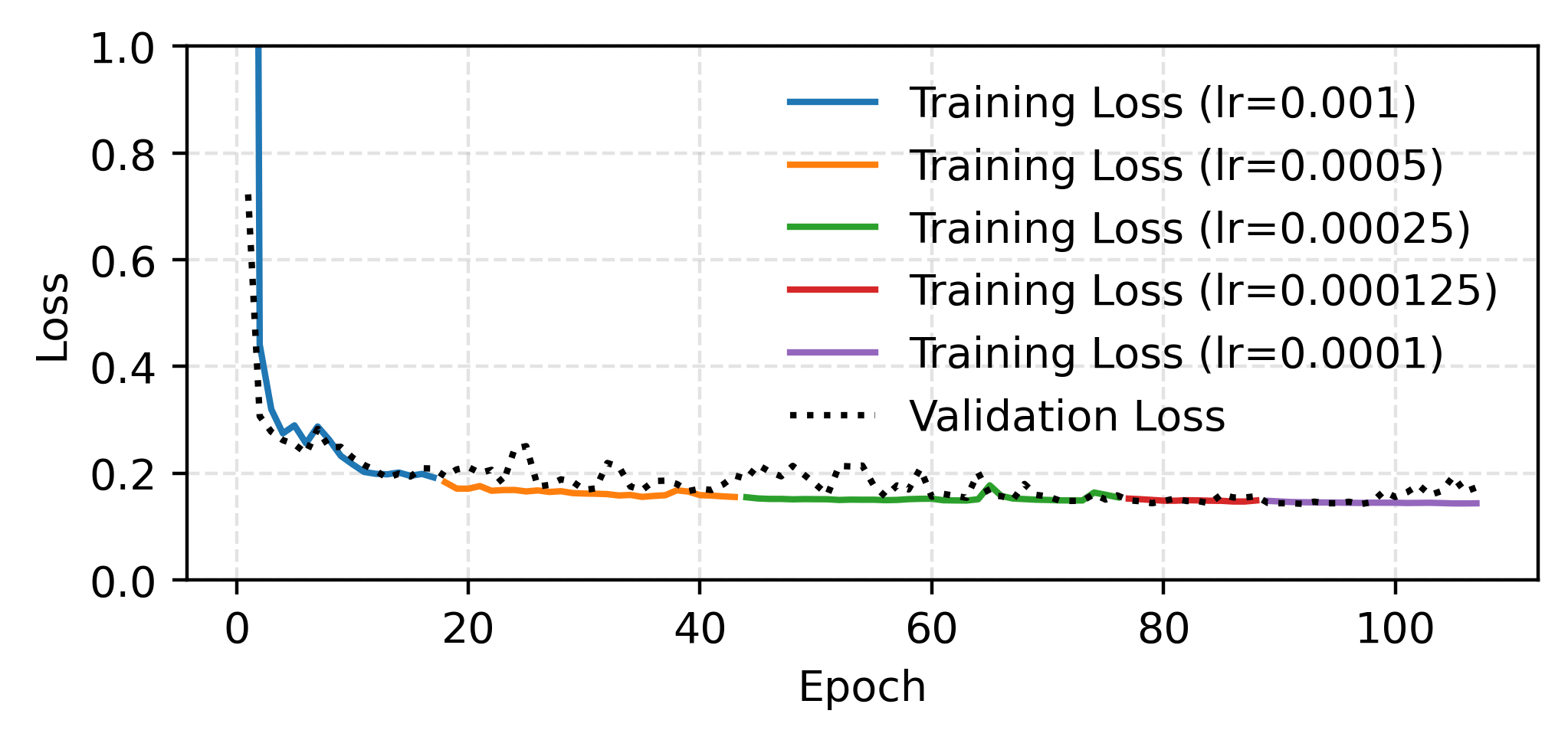}
		\caption{Training and validation loss on the AWGN channel.}
		\label{train_loss}
	\end{figure}
	
	\begin{figure}[t]
		\centering
		\includegraphics[width=1\linewidth]{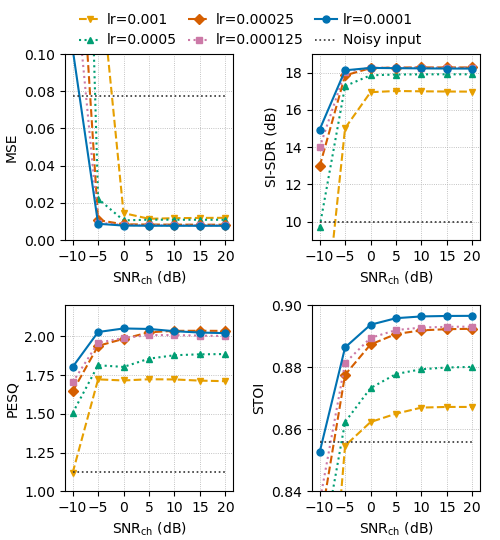}
		\caption{Performance metrics versus channel SNR on the AWGN channel under varying learning rates.}
		\label{lr_metric}
	\end{figure}
	
	Fig.~\ref{train_loss} presents the training and validation loss of the proposed S-DSA on the AWGN channel. 
	
	At the beginning of training, with an initial learning rate of 0.001, the loss decreases rapidly, demonstrating fast convergence in the early stage. As the improvement slows down, the adaptive scheduler automatically reduces the learning rate to 0.0005, under which both training and validation loss continue to decline smoothly with reduced oscillation. When the rate is further decreased to 0.00025 and below, the losses gradually converge and eventually stabilize near 0.14. This progression indicates that large learning rates facilitate fast initial optimization, while smaller rates are essential for fine-tuning and stable convergence. 
	
	Mechanistically, a larger learning rate promotes wide-basin exploration and rapid alignment of feature statistics, accelerating early descent. As the iterate moves into a flatter region of the objective, decaying the rate reduces stochastic gradient variance and prevents overshooting, which explains the diminished oscillation and smooth convergence. Overall, the adaptive learning rate strategy ensures that the proposed S-DSA achieves both rapid convergence and reliable optimization in noisy environments, providing a solid foundation for performance evaluation.

	\subsubsection{Learning-Rate Sensitivity}
	
	Fig.~\ref{lr_metric} shows the performance comparison of the proposed S-DSA under different learning rates across AWGN channel. 
	
	The results show that models trained with a large learning rate of 0.001 achieve only limited enhancement, as reflected by higher MSE, lower SI-SDR, and degraded perceptual scores. When the learning rate is reduced to 0.0005, the performance improves noticeably, yet the enhancement remains unstable across different channel SNR levels. Further decreasing the rate to 0.00025 yields consistently better results, with lower reconstruction error and higher PESQ and STOI values. The optimal performance is observed at 0.000125 and 0.0001, where the proposed S-DSA achieves the lowest MSE, the highest SI-SDR, and superior perceptual quality. At the channel SNR of 0 dB, the proposed S-DSA trained with 0.00025 achieves an SI-SDR of 18.24 dB and a PESQ of 1.98, both higher than those obtained at 0.001 and 0.0005. At 20 dB, the proposed S-DSA trained with 0.0001 further improves the SI-SDR to 18.22 dB and the PESQ to 2.02, confirming the advantage of fine-tuned rates in high-SNR conditions. 
	
	A larger learning rate accelerates the initial descent but tends to compromise generalization, whereas progressively smaller rates allow the proposed S-DSA to refine its parameters and maximize performance. This demonstrates that the adaptive learning rate strategy not only stabilizes training but also ensures robust performance across different channels.
	
	\section{Conclusion}
	\vspace{-1 mm}
	This paper has proposed a semantic communication system, named S-DSA, to address the issues of insufficient semantic information extraction and inadequate noise suppression ability in noisy environments. Experimental results show that, compared with benchmarks, the proposed S-DSA significantly improves perceptual speech quality under low channel SNR conditions while reducing computational overhead. By constructing an E2E architecture, orthogonal decoupling and lightweight reconstruction of semantic–noise features are achieved. The proposed S-DSA achieves efficient compression of semantic features and anti-interference transmission through E2E joint training. This paper provides a cost-effective solution for extracting and transmitting speech semantics, enabling intelligent communication in noisy environments. In future work, we will enhance the model’s robustness against multiple noise types and improve its intelligent adaptation capability.

	
	\bibliographystyle{IEEEtrans}   
	\bibliography{Ref-SDSA-fixed}
	
	\vspace{-15 mm} 
	
	\begin{IEEEbiography}[{\includegraphics[width=1.1in,height=1.4in,clip,keepaspectratio]{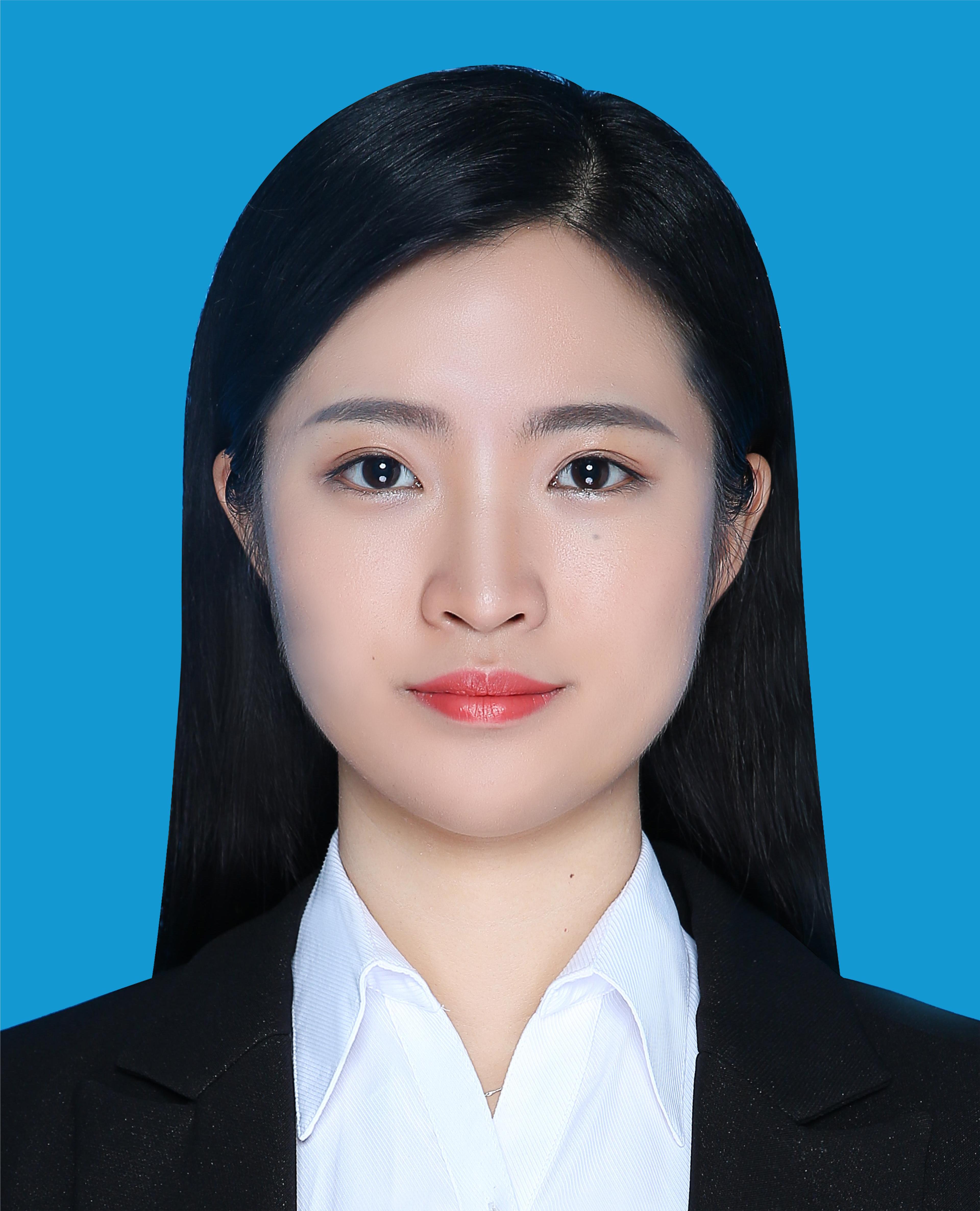}}]{Xiaoling Han}
		(Student Member, IEEE) received the B.S. degree in Electronic Information Engineering from Dalian Maritime University, Dalian, China, in 2020. She is currently working toward the Ph.D. degree at the Department of Information Science and Technology, Dalian Maritime University, Dalian, China, under the supervision of Prof. B. Lin. Her research interests include maritime communications, UAV-based communications and semantic communications.
	\end{IEEEbiography}
	\vspace{-15 mm} 
	\begin{IEEEbiography}[{\includegraphics[width=1.1in,height=1.4in,clip,keepaspectratio]{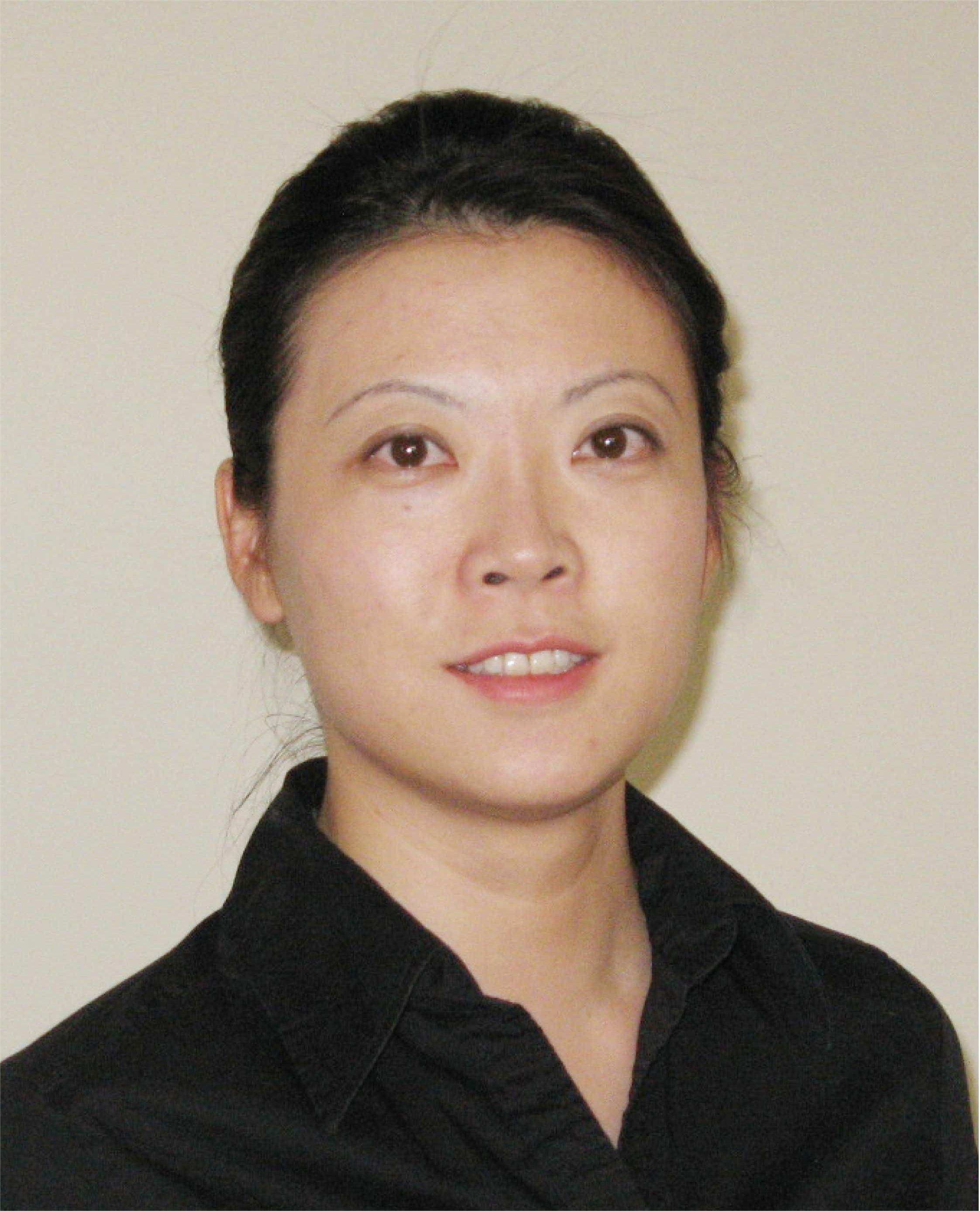}}]{Bin Lin}
		
		(Senior Member, IEEE) received the B.S. and M.S. degrees from Dalian Maritime University, Dalian, China, in 1999 and 2003 respectively, and the Ph.D. degree from the Broadband Communications Research Group, Department of Electrical and Computer Engineering, University of Waterloo, Waterloo, ON, Canada, in 2009. She is currently a Full Professor and Dean of Communication Engineering Department, School of Information Science and Technology, Dalian Maritime University. She has been a Visiting Scholar with George Washington University, Washington, DC, USA, from 2015 to 2016. Her current research interests include wireless communications, network dimensioning and optimization, resource allocation, artificial intelligence, maritime communication networks, edge/cloud computing, wireless sensor networks, and Internet of Things. Dr. Lin is an associate editor of IEEE Transactions on Vehicular Technology and IEEE Internet of Things Journal.
		
	\end{IEEEbiography}
	\vspace{-15 mm}
	\begin{IEEEbiography}[{\includegraphics[width=1.1in,height=1.4in,clip,keepaspectratio]{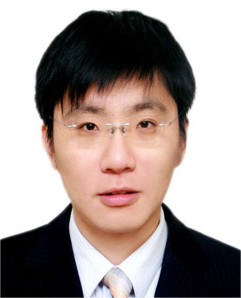}}]{Nan Wu}
		
		(Member, IEEE) received the B.S. degree in electrical and electronics engineering from Dalian University of Technology, Dalian, China, in 2003 and the M.Sc. and Ph.D. degrees from the University of Southampton, Southampton, U.K., in 2004 and 2008, respectively. From 2008 to 2009, he was a Guest Researcher with the National Institute of Standards and Technology, Gaithersburg, MD, USA. Since 2010, he has been an Associate Professor with the School of Electrical and Information Engineering, Dalian Maritime University, Dalian. He has authored one book, more than 50 articles. He also holds two patents. His research interests include deep-learning-based communications, visible light communication, indoor positioning and MIMO.
		
	\end{IEEEbiography}
	\vspace{-15 mm}
	\begin{IEEEbiography}[{\includegraphics[width=1.1in,height=1.4in,clip,keepaspectratio]{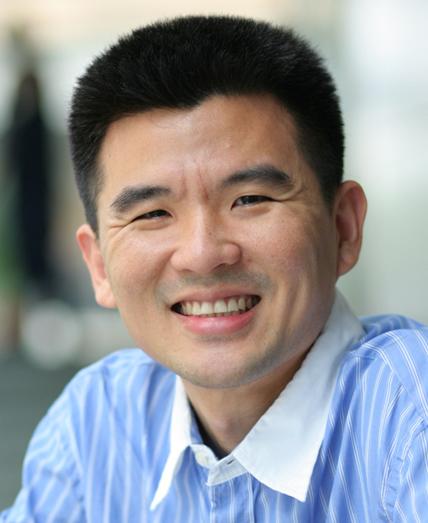}}]{Tony Q. S. Quek}
		
		(S'98-M'08-SM'12-F'18) received the B.E.\ and M.E.\ degrees in electrical and electronics engineering from the Tokyo Institute of Technology in 1998 and 2000, respectively, and the Ph.D.\ degree in electrical engineering and computer science from the Massachusetts Institute of Technology in 2008. Currently, he is the Associate Provost (AI \& Digital Innovation) and Cheng Tsang Man Chair Professor with Singapore University of Technology and Design (SUTD). He also serves as the Director of the Future Communications R\&D Programme, and the ST Engineering Distinguished Professor. He is a co-founder of Silence Laboratories and NeuroRAN. His current research topics include wireless communications and networking, network intelligence, non-terrestrial networks, open radio access network, AI-RAN, and 6G.
		
		Dr.\ Quek was honored with the 2008 Philip Yeo Prize for Outstanding Achievement in Research, the 2012 IEEE William R. Bennett Prize, the 2015 SUTD Outstanding Education Awards -- Excellence in Research, the 2016 IEEE Signal Processing Society Young Author Best Paper Award, the 2017 CTTC Early Achievement Award, the 2017 IEEE ComSoc AP Outstanding Paper Award, the 2020 IEEE Communications Society Young Author Best Paper Award, the 2020 IEEE Stephen O. Rice Prize, the 2020 Nokia Visiting Professor, the 2022 IEEE Signal Processing Society Best Paper Award, the 2024 IIT Bombay International Award For Excellence in Research in Engineering and Technology, the IEEE Communications Society WTC Recognition Award 2024, and the Public Administration Medal (Bronze). He is an IEEE Fellow, a WWRF Fellow, an AIIA Fellow, and a Fellow of the Academy of Engineering Singapore.
	\end{IEEEbiography}
	
\end{document}